\documentclass[aps, pra, showpacs,superscriptaddress , unsortedaddress, twocolumn,
preprintnumbers] {revtex4}
\usepackage{graphicx}
\usepackage{calc}
\usepackage{amsmath}
\usepackage{amssymb}

\newcommand{\D}{\mbox{\rm d}}
\newcommand{\Tr}{\mbox{\rm Tr}}

\newcommand{\ket}[1]{\left|\mathrm{#1}\right\rangle}
\newcommand{\bra}[1]{\left\langle\mathrm{#1}\right|}

\begin{document}
\preprint{PHYSICAL REVIEW A {\bf 81}, 023835 (2010); {\bf 85},
019908(E) (2012)}
\author{A. A. Semenov\footnote{Also at Bogolyubov Institute
for Theoretical Physics, National Academy of Sciences of Ukraine;
sem@iop.kiev.ua.}}

\affiliation{Institute of Physics, National Academy of Sciences of
Ukraine, Prospect Nauky 46, UA-03028 Kiev, Ukraine}%

\author{W. Vogel}
\affiliation{Institut f\"ur Physik, Universit\"{a}t Rostock,
Universit\"{a}tsplatz 3, D-18051 Rostock, Germany}

\title{Entanglement transfer through the turbulent atmosphere}

\begin{abstract}
The propagation of polarization-entangled states of light through
fluctuating loss channels in the turbulent atmosphere is studied,
including the situation of strong losses. We consider violations of
Bell inequalities by light, emitted by a parametric down-conversion
source, after transmission through the turbulent atmosphere. It is
shown by analytical calculations, that in the presence of background
radiation and dark counts, fluctuating loss channels may preserve
entanglement properties of light even better than standard loss
channels, when postselected measurements are applied.
\end{abstract}

\maketitle

\section{Introduction}

The distribution of quantum light through the turbulent atmosphere
has attracted a great deal of attention since recent experiments
\cite{Fedrizzi, Ursin, QKD, Elser} have demonstrated the feasibility
of quantum key distribution in free-space channels. In this
connection the question arises as to whether nonclassical properties
of light can be preserved during its propagation in fluctuating
media. Particularly, special interest is applied to the transfer of
entanglement~\cite{Horodecki} through the turbulent atmosphere since
this problem has important perspectives in quantum communications.

The theory of classical light distributed through the atmosphere was
established many years ago~\cite{Tatarskii-Fante}. Phenomena such as
beam wander, scintillations, beam spreading, and spatial coherence
degradation have been explained in the framework of this
description. It is based on Kolmogorov's theory of turbulence, which
is successfully applied to a description of wave propagation in
random media.

The theory of nonclassical phenomena, such as entanglement, for
light propagation in random media is less developed. First, one
should mention the approach proposed in
Ref.~\cite{DiamentPerinaMilonni} for the description of the
photocounting statistics of quantum light propagating through the
turbulent atmosphere. The idea of this approach is based on the
introduction of a stochastic intensity modulation for light
propagating through random media. Another approach describing
nonclassical properties of light is based on the photon wave
function~\cite{Paterson}. It enables one to describe a special class
of entangled states of light in the turbulent
atmosphere~\cite{Smith}. The concept of the single-photon wave
function can be formulated in a more sophisticated way in terms of
the single-photon Wigner function and then generalized to the
description of arbitrary quantum states. This technique has been
developed in~\cite{Chumak} and applied to some interesting examples
of quantum and classical light.

Recently, a theoretical model for the light distributed through the
turbulent atmosphere and processed by homodyne detection has been
proposed \cite{Semenov1}. The idea is based on the description of
random media in terms of a fluctuating loss channel. It has been
shown that the turbulent atmosphere introduces, compared with
standard loss channels, additional noise into quantum states of
light. The presence of such noise has also been discussed recently
in Ref.~\cite{Marquardt}. Moreover, as has been shown in
Ref.~\cite{Semenov1}, the nonclassicality of bright light fields
(especially fields with a large coherent amplitude) is more fragile
against turbulence than the nonclassicality of weak light. In this
connection special interest should be paid to entangled photon pairs
and single photons, whose fields are weak enough to preserve their
nonclassical properties.

In this paper we give a systematic theoretical description of
polarization-entangled states of light distributed through the
fluctuating loss channels in the atmosphere and processed by
polarization analyzers. For such light we check violations of Bell
inequalities~\cite{Bell, CHSH}. Besides turbulence our consideration
includes imperfect detection and noise counts originating from
internal dark counts and background radiation. Moreover, we deal
with a realistic parametric down-conversion (PDC) source of the
radiation~\cite{Fedrizzi,Ursin, PDC}. We apply the results of our
consideration to analyzing a recently reported
experiment~\cite{Fedrizzi}, where light has been distributed over a
144-km atmospheric channel between two Canary Islands in a
configuration with both receivers being co-located at the same
place. We show that with fluctuating loss channels one may observe
larger values of the Bell parameter compared with corresponding
nonfluctuating loss channels. This unexpected effect is caused by
those (postselected) random  events which are related to small
losses.

The paper is organized as follows. In Sec.~\ref{SecBellExp} we
summarize the needed basic principles of Bell-type experiments. A
systematic description of fluctuating loss channels for the case of
such experiments is given in Sec.~\ref{SecFluctLossChann}. In
Sec.~\ref{SecBellStates} we consider the violation of Bell
inequalities by light after propagation through the turbulent
atmosphere, when the source generates perfect Bell states. The
situation for a realistic PDC source is analyzed in
Sec.~\ref{SecPDC}. In Sec.~\ref{SecMeas} we describe a procedure for
measuring the needed turbulence parameters. A summary and some
concluding remarks are given in Sec.~\ref{SecConcl}.

\section{Bell-type experiments}
\label{SecBellExp}

\begin{figure}[ht!]
\includegraphics[clip=,width=\linewidth]{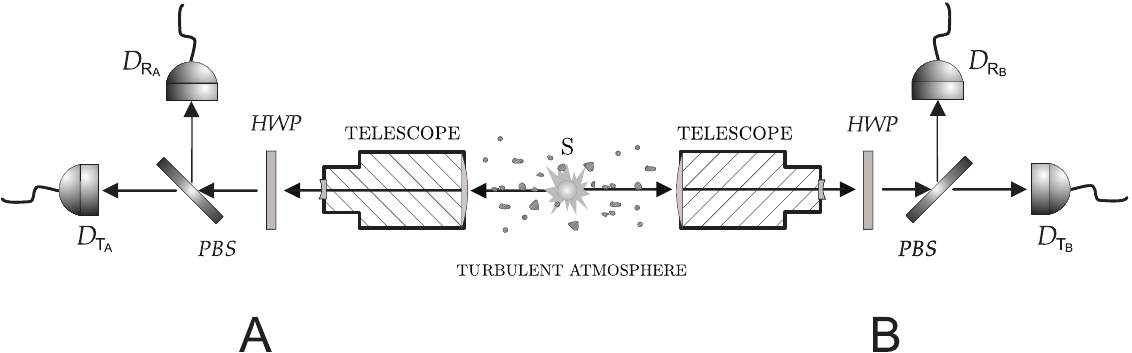}
\caption{\label{Fig1} A typical experimental setup for checking the
violation of Bell inequalities for light transmitting through
turbulent atmosphere. The source $S$ produces entangled photon
pairs. The polarization analyzers at $\mathrm{A}$ and $\mathrm{B}$
sites consist of half-wave plates, $HWP$, polarizer beam-splitters,
$PBS$, and two pairs of detectors: $D_\mathrm{T_A}$ for the
transmitted signal and $D_\mathrm{R_A}$ for the reflected signal at
the $A$ site; $D_\mathrm{T_B}$ for the transmitted signal and
$D_\mathrm{R_B}$ for the reflected signal at the $B$ site.}
\end{figure}

Let us start with the consideration of a typical experimental setup
(see Fig.~\ref{Fig1}). The entangled photon pairs are emitted by the
source $S$ and then transferred through the turbulent atmosphere to
the receiver stations $A$ and $B$. In principle, different
configurations of the considered experiment are possible: The source
can be placed separately from both receivers (e.g., on a low-orbit
satellite), it can be placed on the site of one of the receivers
\cite{Ursin}, or both receivers can be situated at the same
place~\cite{Fedrizzi} (a configuration convenient for testing the
feasibility of entanglement transfer). At the receiver stations the
light is collected by a telescope (or other device) and subsequently
directed to the polarization analyzers. Each of these analyzers
consists of a half-wave plate (which turns the polarization
direction by the angles $\theta_\mathrm{A}$ and $\theta_\mathrm{B}$
at the receiver stations $A$ and $B$, respectively), a polarizing
beam-splitter, and two detectors.

According to photodetection theory~\cite{PhotoDetection,
MandelBook}, the probability of registering one count at the
detector $i_\mathrm{A}=\{\mathrm{T_A},\mathrm{R_A}\}$ by receiver
$A$, one count at the detector
$i_\mathrm{B}=\{\mathrm{T_B},\mathrm{R_B}\}$ by receiver $B$, and no
counts at the other detectors is given by
\begin{equation}
P_\mathrm{i_\mathrm{A}, i_\mathrm{B}}\left(\theta_\mathrm{A},
\theta_\mathrm{B}\right)=\Tr\left(
\hat{\Pi}_{i_\mathrm{A}}^{(1)}\hat{\Pi}_{i_\mathrm{B}}^{(1)}
\hat{\Pi}_{j_\mathrm{A}}^{(0)}\hat{\Pi}_{j_\mathrm{B}}^{(0)}\hat\varrho\right),
\label{Probability1}
\end{equation}
$i_\mathrm{A}\neq j_\mathrm{A}$, $i_\mathrm{B}\neq j_\mathrm{B}$,
where $\hat{\varrho}$ is the density operator of the light field at
the inputs of both polarization analyzers,
\begin{equation}
\hat{\Pi}_{i_\mathrm{A(B)}}^{(n)}=:\frac{\left(\eta_{i_\mathrm{A(B)}}\!
\hat{n}_{i_\mathrm{A(B)}}+N_{i_\mathrm{A(B)}}\right)^n}{n!}
e^{-\eta_{i_\mathrm{A(B)}}\!\hat{n}_{i_\mathrm{A(B)}}-N_{i_\mathrm{A(B)}}}:\label{POVM}
\end{equation}
is the positive operator-valued measure for the detector
$i_\mathrm{A(B)}$~\cite{Semenov3}, $\eta_{i_\mathrm{A(B)}}$ and
$N_{i_\mathrm{A(B)}}$ are the efficiency and the mean value of noise
counts (originating from internal dark counts and background
radiation), respectively, for the detector $i_\mathrm{A(B)}$, and
$:\hspace{0.5em}:$ means normal ordering. The photon number operator
at the input of the detector $i_\mathrm{A(B)}$ can be written in
terms of the corresponding annihilation and creation operators,
\begin{equation}
\hat{n}_{i_\mathrm{A(B)}}=\hat{a}^\dag_{i_\mathrm{A(B)}}\hat{a}_{i_\mathrm{A(B)}}.\label{PhotonNumber}
\end{equation}
These operators can be expressed via the operators of the horizontal
and vertical modes, $\hat{a}_{\scriptscriptstyle \mathrm{H_{A(B)}}}$
and $\hat{a}_{\scriptscriptstyle \mathrm{V_{A(B)}}}$, respectively,
at the input of the polarization analyzers as
\begin{eqnarray}
\hat{a}_{\scriptscriptstyle
T_\mathrm{A(B)}}=\hat{a}_{\scriptscriptstyle
\mathrm{H_{A(B)}}}\cos\theta_\mathrm{A(B)}+
\hat{a}_{\scriptscriptstyle \mathrm{V_{A(B)}}}\sin\theta_\mathrm{A(B)}\label{IOR1},\\
\hat{a}_{\scriptscriptstyle
R_\mathrm{A(B)}}=-\hat{a}_{\scriptscriptstyle
\mathrm{H_{A(B)}}}\sin\theta_\mathrm{A(B)}+
\hat{a}_{\scriptscriptstyle
\mathrm{V_{A(B)}}}\cos\theta_\mathrm{A(B)}\label{IOR2}.
\end{eqnarray}
In the case of using on/off detectors, which cannot distinguish
between different photon numbers, Eq.~(\ref{Probability1}) should be
rewritten as
\begin{equation}
P_\mathrm{i_\mathrm{A}, i_\mathrm{B}}\left(\theta_\mathrm{A},
\theta_\mathrm{B}\right)=\sum\limits_{n,m=1}^{+\infty}\Tr\left(
\hat{\Pi}_{i_\mathrm{A}}^{(n)}\hat{\Pi}_{i_\mathrm{B}}^{(m)}
\hat{\Pi}_{j_\mathrm{A}}^{(0)}\hat{\Pi}_{j_\mathrm{B}}^{(0)}\hat\varrho\right).
\label{Probability}
\end{equation}

The correlation coefficient, which appears in the Bell
theory~\cite{Bell}, is given by
\begin{equation}
E\left(\theta_\mathrm{A}, \theta_\mathrm{B}\right) =
\frac{P_\mathrm{same}\left(\theta_\mathrm{A},
\theta_\mathrm{B}\right)-P_\mathrm{different}\left(\theta_\mathrm{A},
\theta_\mathrm{B}\right)}{P_\mathrm{same}\left(\theta_\mathrm{A},
\theta_\mathrm{B}\right)+P_\mathrm{different}\left(\theta_\mathrm{A},
\theta_\mathrm{B}\right)},\label{correlation}
\end{equation}
where
\begin{equation}
P_\mathrm{same}\left(\theta_\mathrm{A},
\theta_\mathrm{B}\right)=P_\mathrm{\mathrm{T_A},
\mathrm{T_B}}\left(\theta_\mathrm{A},
\theta_\mathrm{B}\right)+P_\mathrm{\mathrm{R_A},
\mathrm{R_B}}\left(\theta_\mathrm{A}, \theta_\mathrm{B}\right)
\end{equation}
is the probability of getting clicks on both detectors in the
transmission channels or both detectors in the reflection channels,
and
\begin{equation}
P_\mathrm{different}\left(\theta_\mathrm{A},
\theta_\mathrm{B}\right)=P_\mathrm{\mathrm{T_A},
\mathrm{R_B}}\left(\theta_\mathrm{A},
\theta_\mathrm{B}\right)+P_\mathrm{\mathrm{R_A},
\mathrm{T_B}}\left(\theta_\mathrm{A}, \theta_\mathrm{B}\right)
\end{equation}
is the probability to get clicks on the detectors in the
transmission channel at one site and the reflection channel at
another site. According to the Clauser-Horne-Shimony-Holt (CHSH)
Bell-type inequality~\cite{CHSH} for any local-realistic theory the
maximal value of the parameter%
\setlength\arraycolsep{0pt}
\begin{eqnarray}
\mathcal{B}&=&\left|E\left(\theta_\mathrm{A}^{(1)},
\theta_\mathrm{B}^{(1)}\right)-E\left(\theta_\mathrm{A}^{(1)},
\theta_\mathrm{B}^{(2)}\right)\right|\label{BellParameter}\\&+&\left|E\left(\theta_\mathrm{A}^{(2)},
\theta_\mathrm{B}^{(2)}\right)+E\left(\theta_\mathrm{A}^{(2)},
\theta_\mathrm{B}^{(1)}\right)\right|\nonumber,
\end{eqnarray}
should be
\begin{equation}
\mathcal{B}\leq 2.
\end{equation}
Quantum light may violate this inequality. In this case the maximal
violating by the value $2\sqrt{2}$ is reached for the Bell state
\begin{eqnarray}
\ket{\mathcal{B}}&=&\frac{1}{\sqrt{2}}\Big(\ket{1}_\mathrm{H_A}\ket{0}_\mathrm{V_A}
\ket{0}_\mathrm{H_B}\ket{1}_\mathrm{V_B}\Big.\label{BellState}\\
\Big.&+&e^{i\varphi} \ket{0}_\mathrm{H_A}\ket{1}_\mathrm{V_A}
\ket{1}_\mathrm{H_B}\ket{0}_\mathrm{V_B}\Big)\nonumber\\
&\equiv&\frac{1}{\sqrt{2}}\Big(\ket{\mathrm{H}}_\mathrm{A}\ket{\mathrm{V}}_\mathrm{B}+
e^{i\varphi}\ket{\mathrm{V}}_\mathrm{A}\ket{\mathrm{H}}_\mathrm{B}\Big),\nonumber
\end{eqnarray}
for $\varphi=\pi$ and $\big(\theta_\mathrm{A}^{(1)},
\theta_\mathrm{B}^{(1)},\theta_\mathrm{A}^{(2)},
\theta_\mathrm{B}^{(2)}\big)=\big(0,\frac{\pi}{8},\frac{\pi}{4},\frac{3\pi}{8}\big)$.

\section{Fluctuating loss channels}
\label{SecFluctLossChann}

The next problem is to derive the density operator $\hat{\varrho}$
for the light transmitted through the turbulent atmosphere.  Similar
to Ref.~\cite{Semenov1}, where homodyne detection of such light has
been studied, we will consider the atmosphere as an attenuating
system. However, an important feature of the present case is that
the nonmonochromatic output mode is not specified by the local
oscillator; it can take an arbitrary form depending on the
distribution of the refraction index in space. Generally speaking,
it is impossible to describe such light in terms of a monochromatic
or nonmonochromatic four-mode density operator. However, it can be
represented in terms of four nonmonochromatic modes with fluctuating
shapes. Of course, such a representation is rather formal since it
is difficult to provide phase-sensitive measurements for such a
mode. However, it appears to be useful for evaluating observable
quantities in the considered experiment.

Let $\hat{a}^\mathrm{in}_{\scriptscriptstyle \mathrm{H_{A(B)}}}$ and
$\hat{a}^\mathrm{in}_{\scriptscriptstyle \mathrm{V_{A(B)}}}$ be the
field annihilation operators of the horizontal and vertical modes,
respectively, generated by the source in the direction of the $A(B)$
receiver station. The operator input--output relations for the
attenuating system can be written as
\begin{eqnarray}
\hat{a}_{\scriptscriptstyle
\mathrm{H_{A(B)}}}=T_\mathrm{H_{A(B)}}\hat{a}^\mathrm{in}_{\scriptscriptstyle
\mathrm{H_{A(B)}}}+T_\mathrm{HV_{A(B)}}\hat{a}^\mathrm{in}_{\scriptscriptstyle
\mathrm{V_{A(B)}}}+R_\mathrm{H_{A(B)}}\hat{c}^\mathrm{in}_{\scriptscriptstyle
\mathrm{H_{A(B)}}},\nonumber \\ \,\label{IOP_O_H1}\\
\hat{a}_{\scriptscriptstyle
\mathrm{V_{A(B)}}}=T_\mathrm{V_{A(B)}}\hat{a}^\mathrm{in}_{\scriptscriptstyle
\mathrm{V_{A(B)}}}+T_\mathrm{VH_{A(B)}}\hat{a}^\mathrm{in}_{\scriptscriptstyle
\mathrm{H_{A(B)}}}+R_\mathrm{V_{A(B)}}\hat{c}^\mathrm{in}_{\scriptscriptstyle
\mathrm{V_{A(B)}}},\nonumber \\ \,\label{IOP_O_V1}
\end{eqnarray}
where $T_\mathrm{H_{A(B)}}$ and $T_\mathrm{V_{A(B)}}$ are the
transmission coefficients for the horizontal and vertical modes,
respectively, in the direction of the $A(B)$ receiver stations;
$\hat{c}^\mathrm{in}_{\scriptscriptstyle \mathrm{H_{A(B)}}}$ and
$\hat{c}^\mathrm{in}_{\scriptscriptstyle \mathrm{V_{A(B)}}}$ are the
operators describing the losses related to absorption and scattering
with the absorption and reflection coefficients
$R_\mathrm{H_{A(B)}}$ and $R_\mathrm{V_{A(B)}}$, respectively. Since
the depolarization effect of the atmosphere is extremely
small~\cite{Tatarskii-Fante}, we may set in the following
$T_\mathrm{HV_{A(B)}}\approx 0$, $T_\mathrm{VH_{A(B)}}\approx 0$,
and $T_\mathrm{H_{A(B)}}\approx T_\mathrm{V_{A(B)}}\equiv
T_\mathrm{A(B)}$. Hence, the operator input--output
relations~(\ref{IOP_O_H1}) and (\ref{IOP_O_V1}) can be simply
written as
\begin{eqnarray}
\hat{a}_{\scriptscriptstyle
\mathrm{H_{A(B)}}}=T_\mathrm{A(B)}\hat{a}^\mathrm{in}_{\scriptscriptstyle
\mathrm{H_{A(B)}}}+R_\mathrm{H_{A(B)}}\hat{c}^\mathrm{in}_{\scriptscriptstyle
\mathrm{H_{A(B)}}},\label{IOP_O_H2}\\
\hat{a}_{\scriptscriptstyle
\mathrm{V_{A(B)}}}=T_\mathrm{A(B)}\hat{a}^\mathrm{in}_{\scriptscriptstyle
\mathrm{V_{A(B)}}}+R_\mathrm{V_{A(B)}}\hat{c}^\mathrm{in}_{\scriptscriptstyle
\mathrm{V_{A(B)}}}.\label{IOP_O_V2}
\end{eqnarray}
The transmission coefficients $T_\mathrm{A(B)}$ and the absorption
and reflection coefficients $R_\mathrm{{H(V)}_{A(B)}}$ satisfy the
constraints
\begin{equation}
\left|T_\mathrm{A(B)}\right|^2+\left|R_\mathrm{{H(V)}_{A(B)}}\right|^2=1,\label{constr1}
\end{equation}

These relations can be converted into the corresponding relations
between the density operator $\hat\varrho_\mathrm{in}$ of the light
generated by the source and the density operator
$\hat{\varrho}_{\bf\scriptscriptstyle T}$ [where ${\bf
T}=\left(T_\mathrm{A},T_\mathrm{B}\right)$] of the light transmitted
through the loss channels under the assumption that the absorption
and scattering system is in the vacuum state. For example, the
relation between the Glauber-Sudarshan $P$
function~\cite{GlauberSudarshan} of the attenuated light,
$P_{\bf\scriptscriptstyle T}\left(\alpha_\mathrm{\scriptscriptstyle
H_{A}},\alpha_\mathrm{\scriptscriptstyle V_{A}},
\alpha_\mathrm{\scriptscriptstyle
H_{B}},\alpha_\mathrm{\scriptscriptstyle V_{B}}\right)$, and the $P$
function of the light generated by the source,
$P_\mathrm{in}\left(\alpha_\mathrm{\scriptscriptstyle
H_{A}},\alpha_\mathrm{\scriptscriptstyle V_{A}},
\alpha_\mathrm{\scriptscriptstyle
H_{B}},\alpha_\mathrm{\scriptscriptstyle V_{B}}\right)$, is given
by~\cite{MandelBook}
\begin{eqnarray}
&&P_{\bf\scriptscriptstyle T}\left(\alpha_\mathrm{\scriptscriptstyle
H_{A}},\alpha_\mathrm{\scriptscriptstyle V_{A}},
\alpha_\mathrm{\scriptscriptstyle
H_{B}},\alpha_\mathrm{\scriptscriptstyle
V_{B}}\right)\nonumber\\
&&=\frac{1}{\left|T_\mathrm{A}\right|^4 \left|T_\mathrm{B}\right|^4}
P_\mathrm{in}\left(\frac{\alpha_\mathrm{\scriptscriptstyle
H_{A}}}{T_\mathrm{A}},\frac{\alpha_\mathrm{\scriptscriptstyle
V_{A}}}{T_\mathrm{A}}, \frac{\alpha_\mathrm{\scriptscriptstyle
H_{B}}}{T_\mathrm{B}},\frac{\alpha_\mathrm{\scriptscriptstyle
V_{B}}}{T_\mathrm{B}}\right). \label{PFuncIOR}
\end{eqnarray}
Similarly, the corresponding characteristic function of the
attenuated light, $\Phi_{\bf\scriptscriptstyle
T}\left(\beta_\mathrm{\scriptscriptstyle
H_{A}},\beta_\mathrm{\scriptscriptstyle V_{A}},
\beta_\mathrm{\scriptscriptstyle
H_{B}},\beta_\mathrm{\scriptscriptstyle V_{B}}\right)$, is related
to the characteristic function of the light generated by the source,
$\Phi_\mathrm{in}\left(\beta_\mathrm{\scriptscriptstyle
H_{A}},\beta_\mathrm{\scriptscriptstyle V_{A}},
\beta_\mathrm{\scriptscriptstyle
H_{B}},\beta_\mathrm{\scriptscriptstyle V_{B}}\right)$, as
\begin{eqnarray}
&&\Phi_{\bf\scriptscriptstyle
T}\left(\beta_\mathrm{\scriptscriptstyle
H_{A}},\beta_\mathrm{\scriptscriptstyle V_{A}},
\beta_\mathrm{\scriptscriptstyle
H_{B}},\beta_\mathrm{\scriptscriptstyle
V_{B}}\right)\label{CharFuncIOR}\\
&&=\Phi_\mathrm{in}\left(T^\ast_\mathrm{A}\beta_\mathrm{\scriptscriptstyle
H_{A}},T^\ast_\mathrm{A}\beta_\mathrm{\scriptscriptstyle V_{A}},
T^\ast_\mathrm{B}\beta_\mathrm{\scriptscriptstyle
H_{B}},T^\ast_\mathrm{B}\beta_\mathrm{\scriptscriptstyle
V_{B}}\right). \nonumber
\end{eqnarray}
The corresponding expressions can also be written for
$s$-parametrized phase-space distribution, for normal-ordered
moments (see, e.g., Ref.~\cite{Semenov1}), for the density operator
in the Fock-state representation~\cite{FockStateEff}, and, in
principle for an arbitrary representation of the density operator.

The main difference of the fluctuating loss channels from the
standard loss channels is that the transmission coefficients
$T_\mathrm{A}$ and $T_\mathrm{B}$ are random variables. This means
that the density operator of the field at the input of the
polarization analyzers, $\hat{\varrho}$, should be obtained by
averaging the density operator $\hat{\varrho}_{\bf\scriptscriptstyle
T}$ with the probability distribution of the transmission
coefficient (PDTC), $\mathcal{P}\left({\mathbf T}\right)$,
\begin{equation}
\hat{\varrho}=\int\limits_{\mathrm{D}}\mathbf{d T}\,
\mathcal{P}\left({\mathbf
T}\right)\,\hat{\varrho}_{\bf\scriptscriptstyle T},\label{IORQS}
\end{equation}
where $\mathbf{d T}=d^2 T_\mathrm{A}\, d^2 T_\mathrm{B}$, and the
integration domain is defined by the conditions
\begin{equation}
\mathrm{D}=\left\{\left|T_\mathrm{A}\right|^2\leq 1,\,
\left|T_\mathrm{B}\right|^2\leq 1\right\}.\label{IntDomain}
\end{equation}
Such an approach is closely related to the idea of a random
intensity modulation in the description of the photocounting
distribution for light transmitted through turbulent
media~\cite{DiamentPerinaMilonni}. An important difference is that
the factor of the intensity modulation was allowed to attain values
in the range of $\left[0,+\infty\right)$. In our approach this
domain is restricted to $\left[0, 1\right]$, which preserves the
required positivity of the density operator in Eq.~(\ref{IORQS}).

\section{Bell States}
\label{SecBellStates}

We start with the consideration of the idealized situation, when the
source generates the Bell state~(\ref{BellState}). This will help us
to better understand the nature of contributions from noise,
atmospheric turbulence, and different experimental imperfections.
First, we consider the density operator (and its entanglement
properties) of such a state after passing the light through
fluctuating loss channels. Next, we include in the consideration
background radiation, dark counts, and detection losses -- for such
conditions we analyze violations of Bell inequalities.

\subsection{Output density operator}

The density operator of the Bell state~(\ref{BellState}) is simply
written as
$\hat\varrho_\mathrm{in}=\hat\varrho_\mathcal{\scriptscriptstyle
B}\equiv\ket{\mathcal{B}}\bra{\mathcal{B}}$. Utilizing
Eqs.~(\ref{PFuncIOR})-(\ref{IORQS}) one gets that the corresponding
density operator for the light at the input ports of the
polarization analyzers is written as a convex combination,
\begin{equation}
\hat{\varrho}=p_{\scriptscriptstyle\mathrm{0}}
\hat{\varrho}_{\scriptscriptstyle\mathrm{0}}+ p_{\scriptscriptstyle
\mathrm{H_{A}}}\hat{\varrho}_{\scriptscriptstyle\mathrm{H_{A}}} +
p_{\scriptscriptstyle
\mathrm{V_{A}}}\hat{\varrho}_{\scriptscriptstyle\mathrm{V_{A}}} +
p_{\scriptscriptstyle
\mathrm{H_{B}}}\hat{\varrho}_{\scriptscriptstyle\mathrm{H_{B}}} +
p_{\scriptscriptstyle
\mathrm{V_{B}}}\hat{\varrho}_{\scriptscriptstyle\mathrm{V_{B}}} +
p_\mathcal{\scriptscriptstyle
B}\hat{\varrho}_\mathcal{\scriptscriptstyle
B},\label{BellStateTurbulence}
\end{equation}
of the following states: vacuum state
$\hat{\varrho}_{\scriptscriptstyle\mathrm{0}}$ with the probability
\begin{equation}
p_{\scriptscriptstyle\mathrm{0}}=\left\langle
\left(1-\left|T_\mathrm{A}\right|^2\right)
\left(1-\left|T_\mathrm{B}\right|^2\right)\right\rangle\Big.\label{p0}
\end{equation}
single-photon states
$\hat{\varrho}_{\scriptscriptstyle\mathrm{H_{A}}}$,
$\hat{\varrho}_{\scriptscriptstyle\mathrm{V_{A}}}$,
$\hat{\varrho}_{\scriptscriptstyle\mathrm{H_{B}}}$, and
$\hat{\varrho}_{\scriptscriptstyle\mathrm{V_{B}}}$ in the
corresponding modes with the probabilities
\begin{equation}
p_{\scriptscriptstyle \mathrm{H_{A}}}=p_{\scriptscriptstyle
\mathrm{V_{A}}}=\frac{1}{2}\left\langle \left|T_\mathrm{A}\right|^2
\left(1-\left|T_\mathrm{B}\right|^2\right)\right\rangle,\label{pHApVA}
\end{equation}
\begin{equation}
p_{\scriptscriptstyle \mathrm{H_{B}}}=p_{\scriptscriptstyle
\mathrm{V_{B}}}=\frac{1}{2}\left\langle \left|T_\mathrm{B}\right|^2
\left(1-\left|T_\mathrm{A}\right|^2\right)\right\rangle,\label{pHBpHA}
\end{equation}
and the Bell state $\hat{\varrho}_{\scriptscriptstyle\mathcal{B}}$
[cf.~Eq.~(\ref{BellState})] with the probability
\begin{equation}
p_{\scriptscriptstyle \mathcal{B}}=\left\langle
\left|T_\mathrm{A}\right|^2\left|T_\mathrm{B}\right|^2\right\rangle.\label{pB}
\end{equation}
The brackets $\left\langle\ldots\right\rangle$ mean averaging with
the PDTC, $\mathcal{P}\left({\mathbf T}\right)$.

In the absence of background radiation and dark counts, only the
contribution of the density operator
$\hat{\varrho}_{\scriptscriptstyle\mathcal{B}}$ in
Eq.~(\ref{BellStateTurbulence}) is postselected in the measurements.
Hence the entanglement properties of the light for such an
experiment are not destroyed by the atmospheric turbulence, at least
in the absence of dark counts, background noise and other
experimental imperfections. The result, nevertheless, may be
significantly changed when such effects play a major role.

\subsection{Imperfect photodetection}

In the presence of background radiation and dark counts, the
postselection procedure no longer results in the perfect separation
of the density matrix
$\hat{\varrho}^{\scriptscriptstyle\mathcal{B}}$ from the
combination~(\ref{BellStateTurbulence}). Indeed, in this case it is
possible that one click at the receiver station $A(B)$ originated
from the signal is combined with a click originated from noise at
the receiver station $B(A)$. Also it is possible that noise
contributes into clicks at both receiver stations.

This situation should be analyzed by substitution of the density
operator~(\ref{BellStateTurbulence}) in Eqs.~(\ref{Probability1})
and (\ref{Probability}) and then the result should be used in
Eq.~(\ref{correlation}). The general analytical solution of this
problem is given in Appendix~\ref{App1}. Here we consider an
important special case of equal detectors and equal background noise
at all detectors, that is, $\eta_{i_\mathrm{A(B)}}=\eta_\mathrm{c}$
and $N_{i_\mathrm{A(B)}}=N_\mathrm{nc}$ for all $i_\mathrm{A(B)}$ in
Eq.~(\ref{POVM}).

After some algebra, the correlation coefficient~(\ref{correlation})
appears to be equal to
\begin{eqnarray}
&&E\left(\theta_\mathrm{A},
\theta_\mathrm{B}\right)\label{CorrelationBell}\\&&=\mathcal{S}\left[-\cos
2\theta_\mathrm{A}\cos 2\theta_\mathrm{B}+\cos\varphi\sin
2\theta_\mathrm{A}\sin 2\theta_\mathrm{B}\right].\nonumber
\end{eqnarray}
For photon-number-resolving detectors we get
\begin{widetext}
\begin{equation}
\mathcal{S}=\frac{p_{\scriptscriptstyle
\mathcal{B}}\eta_\mathrm{c}^2}
{p_{\scriptscriptstyle
\mathcal{B}}\left[\eta_\mathrm{c}+2N_\mathrm{nc}\left(1-\eta_\mathrm{c}\right)\right]^2+
2p_{\scriptscriptstyle
1}N_\mathrm{nc}\left[\eta_\mathrm{c}+2N_\mathrm{nc}\left(1-\eta_\mathrm{c}\right)\right]
+4p_{\scriptscriptstyle\mathrm{0}} N_\mathrm{nc}^2},\label{S_PNR}
\end{equation}
and
\begin{equation}
\mathcal{S}=\frac{p_{\scriptscriptstyle
\mathcal{B}}\eta_\mathrm{c}^2e^{2N_\mathrm{nc}}}
{p_{\scriptscriptstyle
\mathcal{B}}\left[\left(1-\eta_\mathrm{c}\right)
\left(e^{N_\mathrm{nc}}-2\right)+e^{N_\mathrm{nc}}\right]^2+
2p_{\scriptscriptstyle 1}\left[e^{N_\mathrm{nc}}-1\right]
\left[\eta_\mathrm{c}e^{N_\mathrm{nc}}+
2\left(e^{N_\mathrm{nc}}-1\right)\left(1-\eta_\mathrm{c}\right)\right]
+4p_{\scriptscriptstyle\mathrm{0}}
\left[e^{N_\mathrm{nc}}-1\right]^2}\label{S_OnOff}
\end{equation}
for on/off detectors. Herein, $p_{\scriptscriptstyle
1}=p_{\scriptscriptstyle \mathrm{H_{A}}}+p_{\scriptscriptstyle
\mathrm{V_{A}}} +p_{\scriptscriptstyle
\mathrm{H_{B}}}+p_{\scriptscriptstyle \mathrm{V_{B}}}$ is the
probability of the appearance of a single-photon state.
\end{widetext}

\begin{figure}[ht!]
\includegraphics[clip=,width=\linewidth]{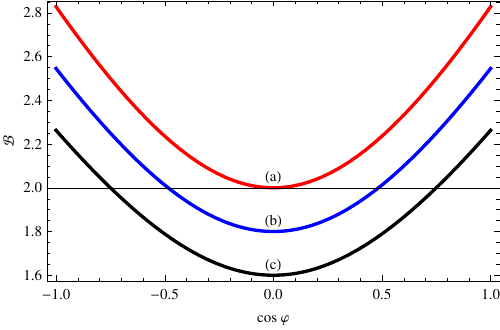}
\caption{\label{Fig2} (Color online) The maximal value of the Bell
parameter $\mathcal{B}$ vs. $\cos\varphi$, for different values of
the parameter $\mathcal{S}$: (a) $1$, (b) $0.9$, (c) $0.8$.}
\end{figure}

The visibility $V=E\left(\frac{\pi}{4},\frac{\pi}{4}\right)$ (see
\cite{Fedrizzi, Ursin}) can be simply obtained from
Eq.~(\ref{CorrelationBell}) as
\begin{equation}
V\left(\varphi\right)=\mathcal{S}\cos\varphi.
\end{equation}
The maximal and minimal values for this parameter are $V_\pm=\pm
\mathcal{S}$. In the case of different detectors and/or different
background radiations at these detectors, see Appendix~\ref{App1},
we obtain $V_{+}\neq -V_{-}$. The parameter $\mathcal{S}$ strictly
depends on both noise counts and turbulence properties of the
atmosphere. Hence, in contrast to the case of perfect
photodetection, the maximal value of the Bell parameter
[Eq.~(\ref{BellParameter})] depends on the atmospheric turbulence.
In Fig.~\ref{Fig2} we consider this parameter as a function of the
parameter $\cos\varphi$, which appears in
Eq.~(\ref{CorrelationBell}).

The obtained analytical results can be used for analyzing a
realistic experimental situation, such as that considered in
Ref.~\cite{Fedrizzi}. In this case both receiver stations had been
positioned at the same place. Consequently, the light pulses at
receivers $A$ and $B$ have been propagating along the same path
through the fluctuating atmosphere. They are separated by a small
time interval, which is much less than the characteristic time of
the atmospheric fluctuations. Hence, in this case one can simply set
\begin{equation}
\left|T_\mathrm{A}\right|^2=
\left|T_\mathrm{B}\right|^2\equiv\eta_\mathrm{atm},\label{ChannelCoorCond}
\end{equation}
where $\eta_\mathrm{atm}$ is the fluctuating atmospheric efficiency.
We also suppose that this efficiency is approximately log-normally
distributed~\cite{Semenov1},
\begin{equation}
\mathcal{P}\left(\eta_\mathrm{atm}\right)=\frac{1}{\sqrt{2\pi}\sigma
\eta_\mathrm{atm}}\exp\left[-\frac{1}{2}\left(\frac{\ln\eta_\mathrm{atm}+
\bar{\theta}}{\sigma}\right)^2\right],\label{LogNormalDistr}
\end{equation}
where $\bar{\theta}=-\left\langle\ln\eta_\mathrm{atm}\right\rangle$
characterizes the mean atmospheric losses and $\sigma$ (the variance
of $\theta=-\ln\eta_\mathrm{atm}$) characterizes the atmosphere
turbulence. It is worth noting that this distribution can be applied
only for $\sigma\ll \bar{\theta}$. Realistic parameters for the
detection efficiency and the mean atmospheric losses can be
extracted from Ref.~\cite{Fedrizzi}: 
$\eta_\mathrm{c}=0.25$, $3\,\mathrm{dB}$ of beam-splitter losses~\cite{BS}, and
$\left\langle T^2\right\rangle=6.3\times 10^{-4}$ corresponding to
$32\,\mathrm{dB}$. In the case of the considered log-normal distribution
$\bar\theta$ is obtained from the relation $\left\langle
T^2\right\rangle{=}e^{-\bar{\theta}+\frac{\sigma^2}{2}}$.

\begin{figure}[ht!]
\includegraphics[clip=]{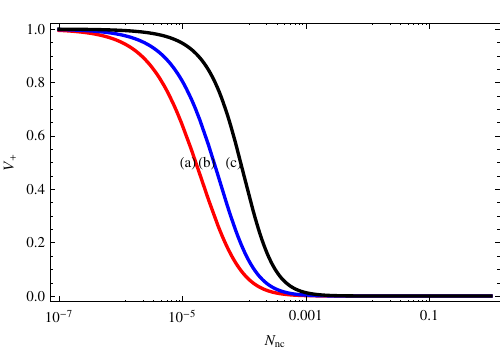}
\caption{\label{Fig3} (Color online) The visibility $V_{+}$ vs the
mean value of noise counts $N_\mathrm{nc}$ for different values of
the turbulence parameter $\sigma$: (a) $0.1$, (b) $1$, (c) $2$. The
detection efficiency is $\eta_\mathrm{c}=0.25$; the mean atmospheric
and beam-splitter losses are $35\,\mathrm{dB}$. The result is equal
for both photon-number-resolving and on/off detectors.}
\end{figure}

An important conclusion is that the visibility in the presence of
loss-fluctuating channels attains higher values compared with
similar standard-loss channels (or with slightly fluctuating loss
channels) (see Fig.~\ref{Fig3}). This fact can be explained by
contributions of random events with $\theta=-\ln\eta_\mathrm{atm}$
(losses) less than the average value $\bar{\theta}$. In the presence
of postselected measurements, this plays a significant role.

\section{Parametric down-conversion source}
\label{SecPDC}

The realistic sources used in experiments generate radiation states,
which are more complicated than the Bell state~(\ref{BellState}).
For example, in Refs.~\cite{Fedrizzi, Ursin} one uses a PDC source
for the generation of entangled photon pairs. For such a source, the
contribution from the multiphoton pair emission is essential. In the
presence of losses or/and in the case of using on/off detectors, all
these photons contribute to the measured Bell parameter. Moreover,
the result appears to be very sensitive to the background noise and
dark counts.

The state generated by the PDC source is given by (cf.~\cite{PDC})
\begin{equation}
\left|\mathrm{PDC}\right\rangle=(\cosh\chi)^{-2}\sum\limits_{n=0}^{+\infty}
\sqrt{n+1}\tanh^n\chi\left|\Phi_n\right\rangle,\label{PDC1}
\end{equation}
where $\chi$ is the squeezing parameter, and
\begin{eqnarray}
&&\left|\Phi_n\right\rangle=\label{PDC2}\\
&&\frac{1}{\sqrt{n+1}}\sum\limits_{m=0}^{n}e^{i\varphi m}
\left|n-m\right\rangle_\mathrm{H_A}\left|m\right\rangle_\mathrm{V_A}
\left|m\right\rangle_\mathrm{H_B}\left|n-m\right\rangle_\mathrm{V_B}\nonumber.
\end{eqnarray}
For $n=1$, $\left|\Phi_n\right\rangle$ is the Bell state,
cf.~Eq.~(\ref{BellState}). The state~(\ref{PDC1}) is Gaussian -- its
characteristic function of the Glauber-Sudarshan $P$ function is
written as
\begin{widetext}
\begin{eqnarray}
\Phi_\mathrm{in}\left(\beta_\mathrm{\scriptscriptstyle
H_{A}},\beta_\mathrm{\scriptscriptstyle V_{A}},
\beta_\mathrm{\scriptscriptstyle
H_{B}},\beta_\mathrm{\scriptscriptstyle
V_{B}}\right)&=&\exp\left[-\frac{\tanh^{2}\chi
\left|\beta_\mathrm{\scriptscriptstyle V_{A}}\right|^2+\tanh^{2}\chi
\left|\beta_\mathrm{\scriptscriptstyle
H_{B}}\right|^2-\tanh\chi\left(\beta_\mathrm{\scriptscriptstyle
V_{A}}\beta_\mathrm{\scriptscriptstyle H_{B}}+
\beta^\ast_\mathrm{\scriptscriptstyle
V_{A}}\beta^\ast_\mathrm{\scriptscriptstyle H_{B}}
\right)}{1-\tanh^{2}\chi}\right]\label{CharFuncPDC}\\
&\times& \exp\left[-\frac{\tanh^{2}\chi
\left|\beta_\mathrm{\scriptscriptstyle H_{A}}\right|^2+\tanh^{2}\chi
\left|\beta_\mathrm{\scriptscriptstyle
V_{B}}\right|^2-\tanh\chi\left(e^{i\varphi}\beta_\mathrm{\scriptscriptstyle
H_{A}}\beta_\mathrm{\scriptscriptstyle V_{B}}+
e^{-i\varphi}\beta^\ast_\mathrm{\scriptscriptstyle
H_{A}}\beta^\ast_\mathrm{\scriptscriptstyle V_{B}}
\right)}{1-\tanh^{2}\chi}\right].\nonumber
\end{eqnarray}
\end{widetext}

We apply Eq.~(\ref{CharFuncPDC}) for the quantum-state input--output
relations~(\ref{CharFuncIOR}) and (\ref{IORQS}). In the resulting
characteristic function, the variables
$\left(\beta_\mathrm{\scriptscriptstyle
H_{A}},\beta_\mathrm{\scriptscriptstyle V_{A}},
\beta_\mathrm{\scriptscriptstyle
H_{B}},\beta_\mathrm{\scriptscriptstyle V_{B}}\right)$ are
transformed to the variables $\left(\beta_\mathrm{\scriptscriptstyle
T_{A}},\beta_\mathrm{\scriptscriptstyle R_{A}},
\beta_\mathrm{\scriptscriptstyle
T_{B}},\beta_\mathrm{\scriptscriptstyle R_{B}}\right)$ using the
input--output relations  for the polarization analyzers~(\ref{IOR1})
and (\ref{IOR2}), which should be rewritten in terms of the
arguments of the characteristic function. Finally, we substitute the
resulting expression in Eqs.~(\ref{Probability}) and
(\ref{Probability1}), which should be also rewritten in terms of the
characteristic function. The obtained result is used in
Eqs.~(\ref{correlation}) and (\ref{BellParameter}) for analyzing
violations of Bell inequalities. For the probability
$P_\mathrm{i_\mathrm{A}, i_\mathrm{B}}\left(\theta_\mathrm{A},
\theta_\mathrm{B}\right)$ we get
\begin{eqnarray}
&&P_\mathrm{i_\mathrm{A}, i_\mathrm{B}}\left(\theta_\mathrm{A},
\theta_\mathrm{B}\right)=\left(1-\tanh^2\chi\right)^4
\exp\left(-\mathbf{N}_\mathrm{nc}\right)
\label{ProbabilitySpecial1}\\
&&\times
\left[2\left\langle\frac{C_\mathrm{i_\mathrm{A}}C_\mathrm{i_\mathrm{B}}}
{C_\mathrm{0}^3}\right\rangle
-\left\langle\frac{C_\mathrm{i_\mathrm{A},i_\mathrm{B}}}{C_\mathrm{0}^2}\right\rangle\right.\nonumber\\
&&- \left.
N_{i_\mathrm{A}}\left\langle\frac{C_\mathrm{i_\mathrm{B}}}{C_\mathrm{0}^2}\right\rangle
-N_{i_\mathrm{B}}\left\langle\frac{C_\mathrm{i_\mathrm{A}}}{C_\mathrm{0}^2}\right\rangle
+N_{i_\mathrm{A}}N_{i_\mathrm{B}}\left\langle\frac{1}{C_\mathrm{0}}\right\rangle
\right]\nonumber
\end{eqnarray}
for photon-number-resolving detectors and
\begin{eqnarray}
&&P_\mathrm{i_\mathrm{A}, i_\mathrm{B}}\left(\theta_\mathrm{A},
\theta_\mathrm{B}\right)=\left(1-\tanh^2\chi\right)^4
\exp\left(-\mathbf{N}_\mathrm{nc}\right)\label{ProbabilitySpecial}\\
&&\times
\left[\left\langle\frac{e^{N_{i_\mathrm{A}}+N_{i_\mathrm{B}}}}{C_\mathrm{0}+C_\mathrm{i_\mathrm{A}}+C_\mathrm{i_\mathrm{B}}
+C_\mathrm{i_\mathrm{A},i_\mathrm{B}}}\right\rangle\right.\nonumber\\&&-
\left.\left\langle\frac{e^{N_{i_\mathrm{A}}}}{C_\mathrm{0}+C_\mathrm{i_\mathrm{A}}}\right\rangle
-\left\langle\frac{e^{N_{i_\mathrm{B}}}}{C_\mathrm{0}+C_\mathrm{i_\mathrm{B}}}\right\rangle
+\left\langle\frac{1}{C_\mathrm{0}}\right\rangle\right]\nonumber
\end{eqnarray}
for on/off detectors. Here
\begin{equation}
\mathbf{N}_\mathrm{nc}=N_{\mathrm{T_A}}+N_{\mathrm{R_A}}+
N_{\mathrm{T_B}}+N_{\mathrm{R_B}}\label{TotalNnc}
\end{equation}
is the total number of noise counts by the four detectors. The
analytical form for the coefficients $C_\mathrm{0}$,
$C_\mathrm{i_\mathrm{A}}$, $C_\mathrm{i_\mathrm{B}}$, and
$C_\mathrm{i_\mathrm{A},i_\mathrm{B}}$ is given in
Appendix~\ref{App2}.

Consider again the case when both receiver stations are situated in
the same place (see~\cite{Fedrizzi}). For simplicity we suppose all
detectors to be equal and all transmission coefficients to be
strongly correlated [i.e. satisfying
condition~(\ref{ChannelCoorCond})]. As follows from the equations in
Appendix~\ref{App2}, in this case one can combine atmospheric and
detection losses in one efficiency,
\begin{equation}
\eta=\eta_\mathrm{atm}\eta_\mathrm{c}.
\end{equation}
We suppose this efficiency to be log-normally distributed, similar
to Eq.~(\ref{LogNormalDistr}). In this case, the coefficients
$C_\mathrm{0}$, $C_\mathrm{i_\mathrm{A}}$,
$C_\mathrm{i_\mathrm{B}}$, and
$C_\mathrm{i_\mathrm{A},i_\mathrm{B}}$ have the form
\begin{equation}
C_\mathrm{0}=\left\{\eta^2\tanh^2\chi-
\left[1+\left(\eta-1\right)\tanh^2\chi\right]^2\right\}^2,\label{C0}
\end{equation}
\begin{eqnarray}
C_\mathrm{i_\mathrm{A}}=C_\mathrm{i_\mathrm{B}}&=&\eta\left(1-\eta\right)\left(1-\tanh^2\chi\right)\tanh^2\chi\label{C1}\\
&\times&\left\{\eta^2\tanh^2\chi-
\left[1+\left(\eta-1\right)\tanh^2\chi\right]^2\right\},\nonumber
\end{eqnarray}
\begin{eqnarray}
C_\mathrm{T_A,T_B}=C_\mathrm{R_A,R_B}&=&\eta^2\tanh^2\chi\left[1-\tanh^2\chi\right]^2\label{Csame}\\
&\times&\left[\left(1-\eta\right)^2 \tanh^2\chi-\sin^2
\left(\theta_\mathrm{A}-\theta_\mathrm{B}\right)\right],\nonumber
\end{eqnarray}
\begin{eqnarray}
C_\mathrm{T_A,R_B}=C_\mathrm{R_A,T_B}&=&
\eta^2\tanh^2\chi\left[1-\tanh^2\chi\right]^2\label{Cdifferent}\\
&\times&\left[\left(1-\eta\right)^2 \tanh^2\chi-\cos^2
\left(\theta_\mathrm{A}-\theta_\mathrm{B}\right)\right].\nonumber
\end{eqnarray}
The realistic values of some parameters can be extracted from
Ref.~\cite{Fedrizzi}. The value of $\bar{\theta}$ is
obtained from the relation $\left\langle
T^2\right\rangle\eta_\mathrm{c}\eta_\mathrm{bs}
{=}e^{-\bar{\theta}+\frac{\sigma^2}{2}}$, where
$\eta_\mathrm{c}=0.25$, $\eta_\mathrm{bs}=0.5$ ($3\,\mathrm{dB}$ of
beam-splitter losses~\cite{BS}), and $\left\langle
T^2\right\rangle=6.3\times 10^{-4}$ ($32\,\mathrm{dB}$ of mean atmospheric
losses). Hence the total single-photon losses are $41\,\mathrm{dB}$. The mean
value of noise counts is $N_\mathrm{nc}=0.5\times10^{-6}$
(200~counts/s from the dark counts and 200~counts/s from the
background radiation in the time window of 1.25~ns).

\begin{figure}[ht!]
\includegraphics[clip=]{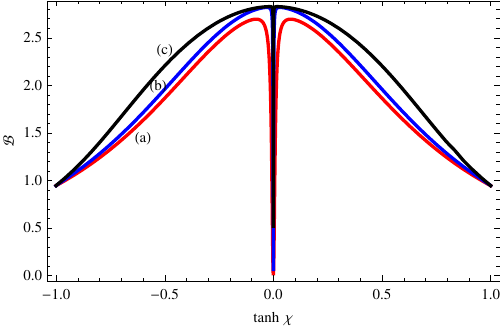}
\caption{\label{Fig4} (Color online) The maximal value of the Bell
parameter $\mathcal{B}$, obtained for photon-number-resolving
detectors, vs the parameter $\tanh\chi$ for the mean value of noise
counts $N_\mathrm{nc}=0.5\times10^{-6}$, the mean total losses
$41\,\mathrm{dB}$, and the turbulence parameter $\sigma$: (a) $0.1$,
(b) $2$, (c) $3$.}
\end{figure}

In the case of photon-number-resolving detectors, the Bell parameter
attains maximal values in the expected range (less than $2\sqrt{2}$
but larger than $2$) for small values of the squeezing parameter
$\chi$ (see Fig.~\ref{Fig4}). The pronounced minimum near
$\tanh\chi=0$ is caused by contributions from classical background
radiation and dark counts. Similar to the case of Bell states, the
fluctuating-loss channels preserve entanglement properties of light
better than similar channels with standard (or slightly fluctuating)
losses. This unexpected property is a consequence of contributions
from random events with $\theta<\bar{\theta}$ in the postselected
measurements, where $\theta=-\ln\eta$.

\begin{figure}[h!]
\includegraphics[clip=]{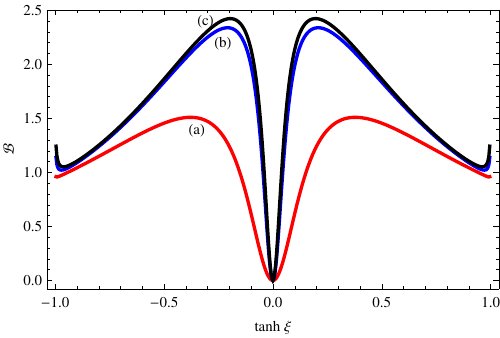}
\caption{\label{Fig5} (Color online) The maximal value of the Bell
parameter $\mathcal{B}$, obtained for on/off detectors, vs the
parameter $\tanh\chi$, for the mean number of noise counts
$N_\mathrm{nc}{=}10^{-5}$, the mean total losses $41\,\mathrm{dB}$,
and the turbulence parameter $\sigma$: (a) $0.1$, (b) $2$, (c) $3$.}
\end{figure}

The situation becomes different in the case of larger
mean numbers of noise counts, e.g. $N_\mathrm{nc}{=}10^{-5}$. We
consider this example for on/off detectors, see Fig.~\ref{Fig5}.
In the case of small fluctuations (or no fluctuations) of the
transmission coefficients, the maximal value of the Bell parameter
may attain values less then 2 for the whole range of
the parameter $\chi$. When the turbulence parameter $\sigma$ becomes
rather large, contributions from events with $\theta<\bar{\theta}$
in the postselected measurements are dominant. This explains the
nonclassical values of the Bell parameter for strong turbulence.

\section{Measurement of turbulence parameters}
\label{SecMeas}

A complete theory of quantum light in the turbulent atmosphere
should include a model for the explicit form of the PDTC. Of course,
the log-normal distribution~(\ref{LogNormalDistr}) cannot be applied
in general. Alternatively, one can consider a procedure that enables
one to reconstruct the PDTC from independent measurements. Such a
procedure has been proposed in Ref.~\cite{Semenov1} for the case of
homodyne detection of the light propagating through the atmosphere.
In the considered case the situation is rather different: Since the
output nonmonochromatic mode has a fluctuating form, it is difficult
to measure its phase properties. On the other hand, the phase
information is not necessary in the considered experiments -- the
corresponding equations (see for example Appendices~\ref{App1} and
\ref{App2}) include only mean values of the functions of
$\left|T_\mathrm{A}\right|^2=\eta_\mathrm{\scriptscriptstyle A}$ and
$\left|T_\mathrm{B}\right|^2=\eta_\mathrm{\scriptscriptstyle B}$.
Hence, one can consider the reconstruction of the PDTC,
$\mathcal{P}\left(\eta_\mathrm{\scriptscriptstyle A},
\eta_\mathrm{\scriptscriptstyle B}\right)$, or alternatively, its
statistical moments $\left\langle\eta_\mathrm{\scriptscriptstyle
A}^n\eta_\mathrm{\scriptscriptstyle B}^m\right\rangle$.

Let the input light be prepared in coherent states,
$\ket{\alpha_\mathrm{\scriptscriptstyle A}}$ in the direction of the
receiver $A$ and $\ket{\alpha_\mathrm{\scriptscriptstyle B}}$ in the
direction of the receiver $B$. For the sake of simplicity we
consider a realistic situation when the resulting signal at the
receivers is much stronger than the background radiation and the dark
counts. In this case the photocounting distribution at the receivers
is given by
\begin{eqnarray}
&&P_{n_\mathrm{A},n_\mathrm{B}}=\int\limits_{0}^{1}\D\eta_\mathrm{\scriptscriptstyle
A}\D\eta_\mathrm{\scriptscriptstyle
B}\mathcal{P}\left(\eta_\mathrm{\scriptscriptstyle A},
\eta_\mathrm{\scriptscriptstyle
B}\right)\label{PhotoDetectionCS}\\
&&\times\frac{\left[\eta_\mathrm{c}\eta_\mathrm{\scriptscriptstyle
A}\left|\alpha_\mathrm{\scriptscriptstyle
A}\right|^2\right]^{n_\mathrm{A}}}{n_\mathrm{A}!}
\frac{\left[\eta_\mathrm{c}\eta_\mathrm{\scriptscriptstyle
B}\left|\alpha_\mathrm{\scriptscriptstyle
B}\right|^2\right]^{n_\mathrm{B}}}{n_\mathrm{B}!}
e^{-\eta_\mathrm{c}\left(\eta_\mathrm{\scriptscriptstyle
A}\left|\alpha_\mathrm{\scriptscriptstyle
A}\right|^2-\eta_\mathrm{\scriptscriptstyle
B}\left|\alpha_\mathrm{\scriptscriptstyle
B}\right|^2\right)},\nonumber
\end{eqnarray}
where $\eta_\mathrm{c}$ is the detection efficiency. In principle,
the methods of ill-posed problems (see, e.g., \cite{Byrne}) enable
one to invert Eq.~(\ref{PhotoDetectionCS}) and to get the PDTC,
$\mathcal{P}\left(\eta_\mathrm{\scriptscriptstyle A},
\eta_\mathrm{\scriptscriptstyle B}\right)$, from the measured
photocounting statistics.

An alternative way to resolve this problem is the reconstruction of
statistical moments of the PDTC. A mean value of a function of
$\eta_\mathrm{\scriptscriptstyle A}$ and
$\eta_\mathrm{\scriptscriptstyle B}$ can be presented as a series of
such moments. These moments can be obtained from the corresponding
moments of photocounts,
\begin{equation}
\overline{n_\mathrm{A}^n
n_\mathrm{B}^m}=\sum\limits_{n_\mathrm{A},n_\mathrm{B}}n_\mathrm{A}^n
n_\mathrm{B}^m\, P_{n_\mathrm{A},n_\mathrm{B}}.
\end{equation}
For example, for the first moments, one gets
\begin{equation}
\overline{n_\mathrm{A}}=\left|\alpha_\mathrm{\scriptscriptstyle
A}\right|^2\left\langle\eta_\mathrm{\scriptscriptstyle
A}\right\rangle,\label{1stMomentA}
\end{equation}
\begin{equation}
\overline{n_\mathrm{B}}=\left|\alpha_\mathrm{\scriptscriptstyle
B}\right|^2\left\langle\eta_\mathrm{\scriptscriptstyle
B}\right\rangle.\label{1stMomentB}
\end{equation}
The first statistical moments of the PDTC,
$\left\langle\eta_\mathrm{\scriptscriptstyle A}\right\rangle$ and
$\left\langle\eta_\mathrm{\scriptscriptstyle B}\right\rangle$, are
easily obtained from these equations. Similarly, the second moments
of photocounts are given by
\begin{equation}
\overline{n_\mathrm{A}n_\mathrm{B}}=\left|\alpha_\mathrm{\scriptscriptstyle
A}\right|^2\left|\alpha_\mathrm{\scriptscriptstyle
B}\right|^2\left\langle\eta_\mathrm{\scriptscriptstyle
A}\eta_\mathrm{\scriptscriptstyle B}\right\rangle,\label{MomentAB}
\end{equation}
\begin{equation}
\overline{n_\mathrm{A}^2}=\left|\alpha_\mathrm{\scriptscriptstyle
A}\right|^4\left\langle\eta_\mathrm{\scriptscriptstyle
A}^2\right\rangle+\left|\alpha_\mathrm{\scriptscriptstyle
A}\right|^2\left\langle\eta_\mathrm{\scriptscriptstyle
A}\right\rangle,\label{2ndMomentA}
\end{equation}
\begin{equation}
\overline{n_\mathrm{B}^2}=\left|\alpha_\mathrm{\scriptscriptstyle
B}\right|^4\left\langle\eta_\mathrm{\scriptscriptstyle
B}^2\right\rangle+\left|\alpha_\mathrm{\scriptscriptstyle
B}\right|^2\left\langle\eta_\mathrm{\scriptscriptstyle
B}\right\rangle.\label{2ndMomentB}
\end{equation}
Combining these equations with Eqs.~(\ref{1stMomentA}) and
(\ref{1stMomentB}), one gets the second statistical moments of the
PDTC from $\overline{n_\mathrm{A}n_\mathrm{B}}$,
$\overline{n_\mathrm{A}^2}$, $\overline{n_\mathrm{B}^2}$,
$\overline{n_\mathrm{A}}$, and $\overline{n_\mathrm{B}}$. In the
same way one can get higher moments of the PDTC by considering the
corresponding moments of photocounts.

\section{Summary and Conclusions}
\label{SecConcl}

In recent experiments violations of Bell inequalities have been
studied, by using entangled  light after transmission through the
turbulent atmosphere. Here we have presented analytical
investigations of the transfer of entanglement through channels with
fluctuating losses. The effect of the atmospheric  turbulence has
been modeled by the statistical properties of the complex
transmission coefficient. We show how the corresponding statistical
characteristics of such channels can be experimentally determined.
Of course, the turbulence itself usually diminishes nonclassical
properties of light. In the considered case only a small
depolarization caused by the atmospheric turbulence might slightly
destroy the entanglement properties of light. However, since this
effect is very small it has been neglected.

For a realistic analysis of  experiments, one must take into account
that the measurements are usually performed in the presence of
background radiation and dark counts. This may also lead, in the
presence of strong losses, to a decrease of the measured value of
the Bell parameter. In this context it is an important result of our
treatment that fluctuating loss channels may preserve entanglement
properties of light even better than the corresponding standard loss
channels. The reason for this fact as follows: Some of the recorded
events are related to small values of the fluctuating atmospheric
transmission. These events are typically caused by background
radiation and dark counts. On the other hand, for strong atmospheric
turbulence the events with randomly occurring large values of the
transmission coefficient will be dominant in the postselected
measurements. In fact, these are the wanted events originating from
the nonclassical light source.

Realistic radiation sources generate quantum states,  which differ
from the perfect Bell states. Important examples are parametric
down-conversion sources, which are often used in experimental
investigations. For such sources we study the situation for both
photon-number-resolving detectors and on/off detectors. The latter
appear to be very sensitive to the presence of background radiation
and dark counts. We have analyzed this problem for realistic
parameters. Unlike in the case of photon-number resolving detectors,
for on/off detectors and for weak turbulence the Bell parameter may
become extremely small. Increasing atmospheric turbulence may
substantially improve this situation.

\acknowledgements

The authors gratefully acknowledge support by the Deutscher
Akademischer Austauschdienst (DAAD) and the Deutsche
Forschungsgemeinschaft (DFG). AAS also thanks the NATO Science for
Peace and Security Programme for financial support.

\begin{widetext}

\appendix

\section{Correlation coefficient for Bell states}
\label{App1}

Here we give the expression for the correlation coefficient
$E\left(\theta_\mathrm{A}, \theta_\mathrm{B}\right)$ for the Bell
state~(12), assuming that all detectors differ from each other. In
the most general case this coefficient is given by%
\setlength\arraycolsep{0pt}
\begin{eqnarray}
&&E\left(\theta_\mathrm{A},
\theta_\mathrm{B}\right)\label{CorrelationBellGen}\\&&=\frac{p_{\scriptscriptstyle
\mathcal{B}}E_{\scriptscriptstyle
\mathcal{B}}\left(\theta_\mathrm{A},
\theta_\mathrm{B}\right)+p_{\scriptscriptstyle
\mathrm{H_{A}}}E_{\scriptscriptstyle
\mathrm{H_{A}}}\left(\theta_\mathrm{A}, \theta_\mathrm{B}\right)+
p_{\scriptscriptstyle \mathrm{V_{A}}}E_{\scriptscriptstyle
\mathrm{V_{A}}}\left(\theta_\mathrm{A}, \theta_\mathrm{B}\right)+
p_{\scriptscriptstyle \mathrm{H_{B}}}E_{\scriptscriptstyle
\mathrm{H_{B}}}\left(\theta_\mathrm{A}, \theta_\mathrm{B}\right)+
p_{\scriptscriptstyle \mathrm{V_{B}}}E_{\scriptscriptstyle
\mathrm{V_{B}}}\left(\theta_\mathrm{A}, \theta_\mathrm{B}\right)
+p_{\scriptscriptstyle \mathrm{0}}E_{\scriptscriptstyle
\mathrm{0}}\left(\theta_\mathrm{A}, \theta_\mathrm{B}\right)
}{p_{\scriptscriptstyle \mathcal{B}}P_{\scriptscriptstyle
\mathcal{B}}\left(\theta_\mathrm{A},
\theta_\mathrm{B}\right)+p_{\scriptscriptstyle
\mathrm{H_{A}}}P_{\scriptscriptstyle
\mathrm{H_{A}}}\left(\theta_\mathrm{A}, \theta_\mathrm{B}\right)+
p_{\scriptscriptstyle \mathrm{V_{A}}}P_{\scriptscriptstyle
\mathrm{V_{A}}}\left(\theta_\mathrm{A}, \theta_\mathrm{B}\right)+
p_{\scriptscriptstyle \mathrm{H_{B}}}P_{\scriptscriptstyle
\mathrm{H_{B}}}\left(\theta_\mathrm{A}, \theta_\mathrm{B}\right)+
p_{\scriptscriptstyle \mathrm{V_{B}}}P_{\scriptscriptstyle
\mathrm{V_{B}}}\left(\theta_\mathrm{A},
\theta_\mathrm{B}\right)+p_{\scriptscriptstyle
\mathrm{0}}P_{\scriptscriptstyle \mathrm{0}}\left(\theta_\mathrm{A},
\theta_\mathrm{B}\right)}.\nonumber
\end{eqnarray}
Here $p_{\scriptscriptstyle \mathrm{0}}$, $p_{\scriptscriptstyle
\mathrm{H_{A}}}$, $p_{\scriptscriptstyle \mathrm{V_{A}}}$,
$p_{\scriptscriptstyle \mathrm{H_{B}}}$, $p_{\scriptscriptstyle
\mathrm{V_{B}}}$, $p_{\scriptscriptstyle \mathcal{B}}$ are given by
Eqs.~(23)-(26). The coefficients $E_{i}\left(\theta_\mathrm{A},
\theta_\mathrm{B}\right)$ and $P_{i}\left(\theta_\mathrm{A},
\theta_\mathrm{B}\right)$ ($i=\left\{0,\mathrm{H_A},\mathrm{V_A},
\mathrm{H_B},\mathrm{V_B},\mathcal{B}\right\}$) have a different
form for the case of photon-number resolving and on/off detectors.
For photon-number resolving detectors they read as
\begin{equation}
E_{\scriptscriptstyle \mathrm{0}}\left(\theta_\mathrm{A},
\theta_\mathrm{B}\right)=e^{-\mathbf{N}_\mathrm{nc}}\left(N_{\scriptscriptstyle
\mathrm{T_{A}}}-N_{\scriptscriptstyle \mathrm{R_{A}}}\right)
\left(N_{\scriptscriptstyle \mathrm{T_{B}}}-N_{\scriptscriptstyle
\mathrm{R_{B}}}\right),
\end{equation}
\begin{eqnarray}\label{EBPNR} E_{\scriptscriptstyle
\mathcal{B}}\left(\theta_\mathrm{A},
\theta_\mathrm{B}\right){=}-\frac{1}{2}e^{-\mathbf{N}_\mathrm{nc}}\Big\{
\left[\eta_{\scriptscriptstyle
\mathrm{T_{A}}}{+}\left(N_{\scriptscriptstyle \mathrm{T_{A}}}{-}
N_{\scriptscriptstyle
\mathrm{R_{A}}}\right)\left(1{-}\eta_{\scriptscriptstyle
\mathrm{T_{A}}}\right)\right]%
\left[\eta_{\scriptscriptstyle
\mathrm{R_{B}}}{+}\left(N_{\scriptscriptstyle \mathrm{R_{B}}}{-}
N_{\scriptscriptstyle
\mathrm{T_{B}}}\right)\left(1{-}\eta_{\scriptscriptstyle
\mathrm{R_{B}}}\right)\right]\Big.\\
\Big.+\left[\eta_{\scriptscriptstyle
\mathrm{R_{A}}}{+}\left(N_{\scriptscriptstyle \mathrm{R_{A}}}{-}
N_{\scriptscriptstyle
\mathrm{T_{A}}}\right)\left(1{-}\eta_{\scriptscriptstyle
\mathrm{R_{A}}}\right)\right]%
\left[\eta_{\scriptscriptstyle
\mathrm{T_{B}}}{+}\left(N_{\scriptscriptstyle \mathrm{T_{B}}}{-}
N_{\scriptscriptstyle
\mathrm{R_{B}}}\right)\left(1{-}\eta_{\scriptscriptstyle
\mathrm{T_{B}}}\right)\right]
\Big\}\nonumber\\
+\frac{1}{2}e^{-\mathbf{N}_\mathrm{nc}}\left[\eta_{\scriptscriptstyle
\mathrm{T_{A}}}+\eta_{\scriptscriptstyle \mathrm{R_{A}}}+
\left(N_{\scriptscriptstyle \mathrm{T_{A}}}{-} N_{\scriptscriptstyle
\mathrm{R_{A}}}\right)\left(\eta_{\scriptscriptstyle
\mathrm{R_{A}}}{-}\eta_{\scriptscriptstyle
\mathrm{T_{A}}}\right)\right]
\left[\eta_{\scriptscriptstyle
\mathrm{T_{B}}}+\eta_{\scriptscriptstyle \mathrm{R_{B}}}+
\left(N_{\scriptscriptstyle \mathrm{T_{B}}}{-} N_{\scriptscriptstyle
\mathrm{R_{B}}}\right)\left(\eta_{\scriptscriptstyle
\mathrm{R_{B}}}{-}\eta_{\scriptscriptstyle
\mathrm{T_{B}}}\right)\right] \nonumber\\
\times\left[\sin^2\left(\theta_\mathrm{A}{+}\theta_\mathrm{B}\right)
\cos^2\frac{\varphi}{2}+
\sin^2\left(\theta_\mathrm{A}{-}\theta_\mathrm{B}\right)
\sin^2\frac{\varphi}{2}\right],\nonumber
\end{eqnarray}
\begin{eqnarray}
E_{\scriptscriptstyle \mathrm{H_{A(B)}}}&&\left(\theta_\mathrm{A},
\theta_\mathrm{B}\right)=e^{-\mathbf{N}_\mathrm{nc}}\left(N_{\scriptscriptstyle
\mathrm{T_{B(A)}}}-N_{\scriptscriptstyle
\mathrm{R_{B(A)}}}\right)\\&&\times\left\{\left[\eta_{\scriptscriptstyle
\mathrm{T_{A(B)}}}+(1-\eta_{\scriptscriptstyle
\mathrm{T_{A(B)}}})(N_{\scriptscriptstyle
\mathrm{T_{A(B)}}}-N_{\scriptscriptstyle
\mathrm{R_{A(B)}}})\right]\cos^{2}\theta_\mathrm{A(B)}-
\left[\eta_{\scriptscriptstyle
\mathrm{R_{A(B)}}}+(1-\eta_{\scriptscriptstyle
\mathrm{R_{A(B)}}})(N_{\scriptscriptstyle
\mathrm{R_{A(B)}}}-N_{\scriptscriptstyle
\mathrm{T_{A(B)}}})\right]\sin^{2}\theta_\mathrm{A(B)}
\right\},\nonumber
\end{eqnarray}
\begin{eqnarray}
E_{\scriptscriptstyle \mathrm{V_{A(B)}}}&&\left(\theta_\mathrm{A},
\theta_\mathrm{B}\right)=e^{-\mathbf{N}_\mathrm{nc}}\left(N_{\scriptscriptstyle
\mathrm{T_{B(A)}}}-N_{\scriptscriptstyle
\mathrm{R_{B(A)}}}\right)\\&&\times\left\{\left[\eta_{\scriptscriptstyle
\mathrm{T_{A(B)}}}+(1-\eta_{\scriptscriptstyle
\mathrm{T_{A(B)}}})(N_{\scriptscriptstyle
\mathrm{T_{A(B)}}}-N_{\scriptscriptstyle
\mathrm{R_{A(B)}}})\right]\sin^{2}\theta_\mathrm{A(B)}-\left[\eta_{\scriptscriptstyle
\mathrm{R_{A(B)}}}+(1-\eta_{\scriptscriptstyle
\mathrm{R_{A(B)}}})(N_{\scriptscriptstyle
\mathrm{R_{A(B)}}}-N_{\scriptscriptstyle
\mathrm{T_{A(B)}}})\right]\cos^{2}\theta_\mathrm{A(B)}
\right\},\nonumber
\end{eqnarray}
\begin{equation}
P_{\scriptscriptstyle \mathrm{0}}\left(\theta_\mathrm{A},
\theta_\mathrm{B}\right)=e^{-\mathbf{N}_\mathrm{nc}}\left(N_{\scriptscriptstyle
\mathrm{T_{A}}}+N_{\scriptscriptstyle \mathrm{R_{A}}}\right)
\left(N_{\scriptscriptstyle \mathrm{T_{B}}}+N_{\scriptscriptstyle
\mathrm{R_{B}}}\right),
\end{equation}
\begin{eqnarray} \label{PBPNR}&&P_{\scriptscriptstyle
\mathcal{B}}\left(\theta_\mathrm{A},
\theta_\mathrm{B}\right)=\frac{1}{2}e^{-\mathbf{N}_\mathrm{nc}}\Big\{
\Big.\left[\eta_{\scriptscriptstyle
\mathrm{T_{A}}}{+}\left(N_{\scriptscriptstyle \mathrm{T_{A}}}{+}
N_{\scriptscriptstyle
\mathrm{R_{A}}}\right)\left(1{-}\eta_{\scriptscriptstyle
\mathrm{T_{A}}}\right)\right]%
\left[\eta_{\scriptscriptstyle
\mathrm{R_{B}}}{+}\left(N_{\scriptscriptstyle \mathrm{T_{B}}}{+}
N_{\scriptscriptstyle
\mathrm{R_{B}}}\right)\left(1{-}\eta_{\scriptscriptstyle
\mathrm{R_{B}}}\right)\right]\Big.\\
&&\Big.+\left[\eta_{\scriptscriptstyle
\mathrm{R_{A}}}{+}\left(N_{\scriptscriptstyle \mathrm{T_{A}}}{+}
N_{\scriptscriptstyle
\mathrm{R_{A}}}\right)\left(1{-}\eta_{\scriptscriptstyle
\mathrm{R_{A}}}\right)\right]%
\left[\eta_{\scriptscriptstyle
\mathrm{T_{B}}}{+}\left(N_{\scriptscriptstyle \mathrm{T_{B}}}{+}
N_{\scriptscriptstyle
\mathrm{R_{B}}}\right)\left(1{-}\eta_{\scriptscriptstyle
\mathrm{T_{B}}}\right)\right]
\Big.\nonumber\\
&&\Big.+\left[\eta_{\scriptscriptstyle
\mathrm{T_{A}}}{-}\eta_{\scriptscriptstyle \mathrm{R_{A}}}\right]
\left[\eta_{\scriptscriptstyle
\mathrm{T_{B}}}{-}\eta_{\scriptscriptstyle \mathrm{R_{B}}}\right]
\left[1{-}N_{\scriptscriptstyle \mathrm{T_{A}}}{-}
N_{\scriptscriptstyle \mathrm{R_{A}}}\right]
\left[1{-}N_{\scriptscriptstyle \mathrm{T_{B}}}{-}
N_{\scriptscriptstyle \mathrm{R_{B}}}\right]
\left[\sin^2\left(\theta_\mathrm{A}{+}\theta_\mathrm{B}\right)
\cos^2\frac{\varphi}{2}+
\sin^2\left(\theta_\mathrm{A}{-}\theta_\mathrm{B}\right)
\sin^2\frac{\varphi}{2}\right] \Big\},\nonumber
\end{eqnarray}
\begin{eqnarray}
P_{\scriptscriptstyle \mathrm{H_{A(B)}}}&&\left(\theta_\mathrm{A},
\theta_\mathrm{B}\right)=e^{-\mathbf{N}_\mathrm{nc}}\left(N_{\scriptscriptstyle
\mathrm{T_{B(A)}}}+N_{\scriptscriptstyle
\mathrm{R_{B(A)}}}\right)\\&&\times\left\{\left[\eta_{\scriptscriptstyle
\mathrm{T_{A(B)}}}+(1-\eta_{\scriptscriptstyle
\mathrm{T_{A(B)}}})(N_{\scriptscriptstyle
\mathrm{T_{A(B)}}}+N_{\scriptscriptstyle
\mathrm{R_{A(B)}}})\right]\cos^{2}\theta_\mathrm{A(B)}+
\left[\eta_{\scriptscriptstyle
\mathrm{R_{A(B)}}}+(1-\eta_{\scriptscriptstyle
\mathrm{R_{A(B)}}})(N_{\scriptscriptstyle
\mathrm{R_{A(B)}}}+N_{\scriptscriptstyle
\mathrm{T_{A(B)}}})\right]\sin^{2}\theta_\mathrm{A(B)}
\right\},\nonumber
\end{eqnarray}
\begin{eqnarray}
P_{\scriptscriptstyle \mathrm{V_{A(B)}}}&&\left(\theta_\mathrm{A},
\theta_\mathrm{B}\right)=e^{-\mathbf{N}_\mathrm{nc}}\left(N_{\scriptscriptstyle
\mathrm{T_{B(A)}}}+N_{\scriptscriptstyle
\mathrm{R_{B(A)}}}\right)\\&&\times\left\{\left[\eta_{\scriptscriptstyle
\mathrm{T_{A(B)}}}+(1-\eta_{\scriptscriptstyle
\mathrm{T_{A(B)}}})(N_{\scriptscriptstyle
\mathrm{T_{A(B)}}}+N_{\scriptscriptstyle
\mathrm{R_{A(B)}}})\right]\sin^{2}\theta_\mathrm{A(B)}+\left[\eta_{\scriptscriptstyle
\mathrm{R_{A(B)}}}+(1-\eta_{\scriptscriptstyle
\mathrm{R_{A(B)}}})(N_{\scriptscriptstyle
\mathrm{R_{A(B)}}}+N_{\scriptscriptstyle
\mathrm{T_{A(B)}}})\right]\cos^{2}\theta_\mathrm{A(B)}
\right\},\nonumber
\end{eqnarray}
where $\mathbf{N}_\mathrm{nc}$ is the total mean value of noise
counts, cf.~Eq.~(38). For the case of using on/off detectors the
coefficients in Eq.~(\ref{CorrelationBellGen}) are given by
\begin{equation}
E_{\scriptscriptstyle \mathrm{0}}\left(\theta_\mathrm{A},
\theta_\mathrm{B}\right)=e^{-\mathbf{N}_\mathrm{nc}}\left(e^{N_{\scriptscriptstyle
\mathrm{T_{A}}}}-e^{N_{\scriptscriptstyle \mathrm{R_{A}}}}\right)
\left(e^{N_{\scriptscriptstyle
\mathrm{T_{B}}}}-e^{N_{\scriptscriptstyle \mathrm{R_{B}}}}\right),
\end{equation}
\begin{eqnarray}\label{EBOnOff} E_{\scriptscriptstyle
\mathcal{B}}\left(\theta_\mathrm{A},
\theta_\mathrm{B}\right){=}-\frac{1}{2}e^{-\mathbf{N}_\mathrm{nc}}\Big\{
\left[\eta_{\scriptscriptstyle
\mathrm{T_{A}}}e^{N_{\scriptscriptstyle
\mathrm{R_{A}}}}{+}e^{N_{\scriptscriptstyle
\mathrm{T_{A}}}}{-}e^{N_{\scriptscriptstyle
\mathrm{R_{A}}}}\right]%
\left[\eta_{\scriptscriptstyle
\mathrm{R_{B}}}e^{N_{\scriptscriptstyle
\mathrm{T_{B}}}}{+}e^{N_{\scriptscriptstyle
\mathrm{R_{B}}}}{-}e^{N_{\scriptscriptstyle
\mathrm{T_{B}}}}\right]\Big.\\
\Big.+ \left[\eta_{\scriptscriptstyle
\mathrm{R_{A}}}e^{N_{\scriptscriptstyle
\mathrm{T_{A}}}}{+}e^{N_{\scriptscriptstyle
\mathrm{R_{A}}}}{-}e^{N_{\scriptscriptstyle
\mathrm{T_{A}}}}\right]%
\left[\eta_{\scriptscriptstyle
\mathrm{T_{B}}}e^{N_{\scriptscriptstyle
\mathrm{R_{B}}}}{+}e^{N_{\scriptscriptstyle
\mathrm{T_{B}}}}{-}e^{N_{\scriptscriptstyle \mathrm{R_{B}}}}\right]
\Big\}\nonumber\\
+\frac{1}{2}e^{-\mathbf{N}_\mathrm{nc}}\left[\eta_{\scriptscriptstyle
\mathrm{T_{A}}}e^{N_{\scriptscriptstyle
\mathrm{R_{A}}}}+\eta_{\scriptscriptstyle
\mathrm{R_{A}}}e^{N_{\scriptscriptstyle \mathrm{T_{A}}}}\right]
\left[\eta_{\scriptscriptstyle
\mathrm{T_{B}}}e^{N_{\scriptscriptstyle
\mathrm{R_{B}}}}+\eta_{\scriptscriptstyle
\mathrm{R_{B}}}e^{N_{\scriptscriptstyle \mathrm{T_{B}}}}\right] \nonumber\\
\times\left[\sin^2\left(\theta_\mathrm{A}{+}\theta_\mathrm{B}\right)
\cos^2\frac{\varphi}{2}+
\sin^2\left(\theta_\mathrm{A}{-}\theta_\mathrm{B}\right)
\sin^2\frac{\varphi}{2}\right],\nonumber
\end{eqnarray}
\begin{eqnarray}
E_{\scriptscriptstyle \mathrm{H_{A(B)}}}\left(\theta_\mathrm{A},
\theta_\mathrm{B}\right)&&=e^{-\mathbf{N}_\mathrm{nc}}\left(e^{N_{\scriptscriptstyle
\mathrm{T_{B(A)}}}}-e^{N_{\scriptscriptstyle
\mathrm{R_{B(A)}}}}\right)\\&&\times\Big\{\left[\eta_{\scriptscriptstyle
\mathrm{T_{A(B)}}}e^{N_{\scriptscriptstyle
\mathrm{T_{A(B)}}}}+(1-\eta_{\scriptscriptstyle
\mathrm{T_{A(B)}}})(e^{N_{\scriptscriptstyle
\mathrm{T_{A(B)}}}}-e^{N_{\scriptscriptstyle
\mathrm{R_{A(B)}}}})\right]\cos^{2}\theta_\mathrm{A(B)}\Big.
\nonumber \\ &&\Big. -\left[\eta_{\scriptscriptstyle
\mathrm{R_{A(B)}}}e^{N_{\scriptscriptstyle
\mathrm{R_{A(B)}}}}+(1-\eta_{\scriptscriptstyle
\mathrm{R_{A(B)}}})(e^{N_{\scriptscriptstyle
\mathrm{R_{A(B)}}}}-e^{N_{\scriptscriptstyle
\mathrm{T_{A(B)}}}})\right]\sin^{2}\theta_\mathrm{A(B)}
\Big\},\nonumber
\end{eqnarray}
\begin{eqnarray}
E_{\scriptscriptstyle \mathrm{V_{A(B)}}}\left(\theta_\mathrm{A},
\theta_\mathrm{B}\right)&&=e^{-\mathbf{N}_\mathrm{nc}}\left(e^{N_{\scriptscriptstyle
\mathrm{T_{B(A)}}}}-e^{N_{\scriptscriptstyle
\mathrm{R_{B(A)}}}}\right)\\&&\times\Big\{\left[\eta_{\scriptscriptstyle
\mathrm{T_{A(B)}}}e^{N_{\scriptscriptstyle
\mathrm{T_{A(B)}}}}+(1-\eta_{\scriptscriptstyle
\mathrm{T_{A(B)}}})(e^{N_{\scriptscriptstyle
\mathrm{T_{A(B)}}}}-e^{N_{\scriptscriptstyle
\mathrm{R_{A(B)}}}})\right]\sin^{2}\theta_\mathrm{A(B)}\Big.
\nonumber
\\ &&\Big. -\left[\eta_{\scriptscriptstyle
\mathrm{R_{A(B)}}}e^{N_{\scriptscriptstyle
\mathrm{R_{A(B)}}}}+(1-\eta_{\scriptscriptstyle
\mathrm{R_{A(B)}}})(e^{N_{\scriptscriptstyle
\mathrm{R_{A(B)}}}}-e^{N_{\scriptscriptstyle
\mathrm{T_{A(B)}}}})\right]\cos^{2}\theta_\mathrm{A(B)}
\Big\},\nonumber
\end{eqnarray}
\begin{equation}
P_{\scriptscriptstyle \mathrm{0}}\left(\theta_\mathrm{A},
\theta_\mathrm{B}\right)=e^{-\mathbf{N}_\mathrm{nc}}\left(e^{N_{\scriptscriptstyle
\mathrm{T_{A}}}}+e^{N_{\scriptscriptstyle \mathrm{R_{A}}}}-2\right)
\left(e^{N_{\scriptscriptstyle
\mathrm{T_{B}}}}+e^{N_{\scriptscriptstyle \mathrm{R_{B}}}}-2\right),
\end{equation}
\begin{eqnarray}\label{PBOnOff} P_{\scriptscriptstyle
\mathcal{B}}\left(\theta_\mathrm{A},
\theta_\mathrm{B}\right)=\frac{1}{2}e^{-\mathbf{N}_\mathrm{nc}}\Big\{
\Big.\left[\left(1{-}\eta_{\scriptscriptstyle
\mathrm{T_{A}}}\right)\left(e^{N_{\scriptscriptstyle\mathrm{R_{A}}}}
-2\right){+}e^{N_{\scriptscriptstyle\mathrm{T_{A}}}} \right]%
\left[\left(1{-}\eta_{\scriptscriptstyle
\mathrm{R_{B}}}\right)\left(e^{N_{\scriptscriptstyle\mathrm{T_{B}}}}
-2\right){+}e^{N_{\scriptscriptstyle\mathrm{R_{B}}}} \right]\Big.\\
\Big.+ \left[\left(1{-}\eta_{\scriptscriptstyle
\mathrm{R_{A}}}\right)\left(e^{N_{\scriptscriptstyle\mathrm{T_{A}}}}
-2\right){+}e^{N_{\scriptscriptstyle\mathrm{R_{A}}}} \right]%
\left[\left(1{-}\eta_{\scriptscriptstyle
\mathrm{T_{B}}}\right)\left(e^{N_{\scriptscriptstyle\mathrm{R_{B}}}}
-2\right){+}e^{N_{\scriptscriptstyle\mathrm{T_{B}}}}\right]
\Big.\nonumber\\
\Big.+\left[2\left(\eta_{\scriptscriptstyle
\mathrm{T_{A}}}{-}\eta_{\scriptscriptstyle
\mathrm{R_{A}}}\right){+}\eta_{\scriptscriptstyle
\mathrm{R_{A}}}e^{N_{\scriptscriptstyle\mathrm{T_{A}}}}{-}
\eta_{\scriptscriptstyle
\mathrm{T_{A}}}e^{N_{\scriptscriptstyle\mathrm{R_{A}}}}\right]
\left[2\left(\eta_{\scriptscriptstyle
\mathrm{T_{B}}}{-}\eta_{\scriptscriptstyle
\mathrm{R_{B}}}\right){+}\eta_{\scriptscriptstyle
\mathrm{R_{B}}}e^{N_{\scriptscriptstyle\mathrm{T_{B}}}}{-}
\eta_{\scriptscriptstyle
\mathrm{T_{B}}}e^{N_{\scriptscriptstyle\mathrm{R_{B}}}}\right]\nonumber\\
\times\left[\sin^2\left(\theta_\mathrm{A}{+}\theta_\mathrm{B}\right)
\cos^2\frac{\varphi}{2}+
\sin^2\left(\theta_\mathrm{A}{-}\theta_\mathrm{B}\right)
\sin^2\frac{\varphi}{2}\right] \Big\},\nonumber
\end{eqnarray}
\begin{eqnarray}
P_{\scriptscriptstyle \mathrm{H_{A(B)}}}\left(\theta_\mathrm{A},
\theta_\mathrm{B}\right)&&=e^{-\mathbf{N}_\mathrm{nc}}\left(e^{N_{\scriptscriptstyle
\mathrm{T_{B(A)}}}}+e^{N_{\scriptscriptstyle
\mathrm{R_{B(A)}}}}-2\right)\\&&\times\Big\{\left[\eta_{\scriptscriptstyle
\mathrm{T_{A(B)}}}e^{N_{\scriptscriptstyle
\mathrm{T_{A(B)}}}}+(1-\eta_{\scriptscriptstyle
\mathrm{T_{A(B)}}})(e^{N_{\scriptscriptstyle
\mathrm{T_{A(B)}}}}+e^{N_{\scriptscriptstyle
\mathrm{R_{A(B)}}}}-2)\right]\cos^{2}\theta_\mathrm{A(B)}\Big.
\nonumber \\ &&\Big. -\left[\eta_{\scriptscriptstyle
\mathrm{R_{A(B)}}}e^{N_{\scriptscriptstyle
\mathrm{R_{A(B)}}}}+(1-\eta_{\scriptscriptstyle
\mathrm{R_{A(B)}}})(e^{N_{\scriptscriptstyle
\mathrm{R_{A(B)}}}}+e^{N_{\scriptscriptstyle
\mathrm{T_{A(B)}}}}-2)\right]\sin^{2}\theta_\mathrm{A(B)}
\Big\},\nonumber
\end{eqnarray}
\begin{eqnarray}
P_{\scriptscriptstyle \mathrm{V_{A(B)}}}\left(\theta_\mathrm{A},
\theta_\mathrm{B}\right)&&=e^{-\mathbf{N}_\mathrm{nc}}\left(e^{N_{\scriptscriptstyle
\mathrm{T_{B(A)}}}}+e^{N_{\scriptscriptstyle
\mathrm{R_{B(A)}}}}-2\right)\\&&\times\Big\{\left[\eta_{\scriptscriptstyle
\mathrm{T_{A(B)}}}e^{N_{\scriptscriptstyle
\mathrm{T_{A(B)}}}}+(1-\eta_{\scriptscriptstyle
\mathrm{T_{A(B)}}})(e^{N_{\scriptscriptstyle
\mathrm{T_{A(B)}}}}+e^{N_{\scriptscriptstyle
\mathrm{R_{A(B)}}}}-2)\right]\sin^{2}\theta_\mathrm{A(B)}\Big.
\nonumber
\\ &&\Big. +\left[\eta_{\scriptscriptstyle
\mathrm{R_{A(B)}}}e^{N_{\scriptscriptstyle
\mathrm{R_{A(B)}}}}+(1-\eta_{\scriptscriptstyle
\mathrm{R_{A(B)}}})(e^{N_{\scriptscriptstyle
\mathrm{R_{A(B)}}}}-e^{N_{\scriptscriptstyle
\mathrm{T_{A(B)}}}}-2)\right]\cos^{2}\theta_\mathrm{A(B)}
\Big\}.\nonumber
\end{eqnarray}

\section{Coefficients in the detection probabilities for the parametric down-conversion source}
\label{App2}

Here we give the explicit form of the coefficients $C_\mathrm{0}$,
$C_\mathrm{i_\mathrm{A}}$, $C_\mathrm{i_\mathrm{B}}$, and
$C_\mathrm{i_\mathrm{A},i_\mathrm{B}}$ in the most general case.
These coefficients appear in the expressions for the probabilities
$P_\mathrm{i_\mathrm{A}, i_\mathrm{B}}\left(\theta_\mathrm{A},
\theta_\mathrm{B}\right)$, cf.~Eqs.~(36) and (37), in the case of
using a PDC source of
entangled photons. They are given by%
\setlength\arraycolsep{0pt}
\begin{eqnarray}
&&C_\mathrm{0}=\left[1+\left(\Omega_\mathrm{T_{A}}-1\right)\tanh^2\chi\right]\left[1+\left(\Omega_\mathrm{R_{A}}-1\right)\tanh^2\chi\right] \left[1+\left(\Omega_\mathrm{T_{B}}-1\right)\tanh^2\chi\right]\left[1+\left(\Omega_\mathrm{R_{B}}-1\right)\tanh^2\chi\right]\\
&&-\left[1+\left(\Omega_\mathrm{T_{A}}-1\right)\tanh^2\chi\right]\left[1+\left(\Omega_\mathrm{T_{B}}-1\right)\tanh^2\chi\right]D_\mathrm{T_A,T_B}
-\left[1+\left(\Omega_\mathrm{T_{A}}-1\right)\tanh^2\chi\right]\left[1+\left(\Omega_\mathrm{R_{B}}-1\right)\tanh^2\chi\right]D_\mathrm{T_A,R_B}\nonumber\\
&&-\left[1+\left(\Omega_\mathrm{R_{A}}-1\right)\tanh^2\chi\right]\left[1+\left(\Omega_\mathrm{R_{B}}-1\right)\tanh^2\chi\right]D_\mathrm{R_A,R_B}
-\left[1+\left(\Omega_\mathrm{R_{A}}-1\right)\tanh^2\chi\right]\left[1+\left(\Omega_\mathrm{T_{B}}-1\right)\tanh^2\chi\right]D_\mathrm{R_A,T_B}\nonumber\\
&&+D_\mathrm{0},\nonumber
\end{eqnarray}
\begin{eqnarray}
&&C_\mathrm{T_{A}}=-\Omega_\mathrm{T_{A}}\left[1+\left(\Omega_\mathrm{R_{A}}-1\right)\tanh^2\chi\right] \left[1+\left(\Omega_\mathrm{T_{B}}-1\right)\tanh^2\chi\right]\left[1+\left(\Omega_\mathrm{R_{B}}-1\right)\tanh^2\chi\right]\\
&&+\Omega_\mathrm{T_{A}}\tanh^2\chi\left[1+\left(\Omega_\mathrm{T_{B}}-1\right)\tanh^2\chi\right]D_\mathrm{T_A,T_B}
+\Omega_\mathrm{T_{A}}\tanh^2\chi\left[1+\left(\Omega_\mathrm{R_{B}}-1\right)\tanh^2\chi\right]D_\mathrm{T_A,R_B}\nonumber\\
&&+\left[1+\left(\Omega_\mathrm{R_{A}}-1\right)\tanh^2\chi\right]\left[1+\left(\Omega_\mathrm{R_{B}}-1\right)\tanh^2\chi\right]D_\mathrm{R_A,R_B}
+\left[1+\left(\Omega_\mathrm{R_{A}}-1\right)\tanh^2\chi\right]\left[1+\left(\Omega_\mathrm{T_{B}}-1\right)\tanh^2\chi\right]D_\mathrm{R_A,T_B}\nonumber\\
&&-D_\mathrm{0},\nonumber
\end{eqnarray}
\begin{eqnarray}
&&C_\mathrm{R_{A}}=-\Omega_\mathrm{R_{A}}\tanh^2\chi\left[1+\left(\Omega_\mathrm{T_{A}}-1\right)\tanh^2\chi\right]\left[1+\left(\Omega_\mathrm{T_{B}}-1\right)\tanh^2\chi\right]\left[1+\left(\Omega_\mathrm{R_{B}}-1\right)\tanh^2\chi\right]\\
&&+\left[1+\left(\Omega_\mathrm{T_{A}}-1\right)\tanh^2\chi\right]\left[1+\left(\Omega_\mathrm{T_{B}}-1\right)\tanh^2\chi\right]D_\mathrm{T_A,T_B}
+\left[1+\left(\Omega_\mathrm{T_{A}}-1\right)\tanh^2\chi\right]\left[1+\left(\Omega_\mathrm{R_{B}}-1\right)\tanh^2\chi\right]D_\mathrm{T_A,R_B}\nonumber\\
&&+\Omega_\mathrm{R_{A}}\tanh^2\chi\left[1+\left(\Omega_\mathrm{R_{B}}-1\right)\tanh^2\chi\right]D_\mathrm{R_A,R_B}
+\Omega_\mathrm{R_{A}}\tanh^2\chi\left[1+\left(\Omega_\mathrm{T_{B}}-1\right)\tanh^2\chi\right]D_\mathrm{R_A,T_B}\nonumber\\
&&-D_\mathrm{0},\nonumber
\end{eqnarray}
\begin{eqnarray}
&&C_\mathrm{T_{B}}=-\Omega_\mathrm{T_{B}}\tanh^2\chi\left[1+\left(\Omega_\mathrm{T_{A}}-1\right)\tanh^2\chi\right]\left[1+\left(\Omega_\mathrm{R_{A}}-1\right)\tanh^2\chi\right]\left[1+\left(\Omega_\mathrm{R_{B}}-1\right)\tanh^2\chi\right]\\
&&+\Omega_\mathrm{T_{B}}\tanh^2\chi\left[1+\left(\Omega_\mathrm{T_{A}}-1\right)\tanh^2\chi\right]D_\mathrm{T_A,T_B}
+\left[1+\left(\Omega_\mathrm{T_{A}}-1\right)\tanh^2\chi\right]\left[1+\left(\Omega_\mathrm{R_{B}}-1\right)\tanh^2\chi\right]D_\mathrm{T_A,R_B}\nonumber\\
&&+\left[1+\left(\Omega_\mathrm{R_{A}}-1\right)\tanh^2\chi\right]\left[1+\left(\Omega_\mathrm{R_{B}}-1\right)\tanh^2\chi\right]D_\mathrm{R_A,R_B}
+\Omega_\mathrm{T_{B}}\tanh^2\chi\left[1+\left(\Omega_\mathrm{R_{A}}-1\right)\tanh^2\chi\right]D_\mathrm{R_A,T_B}\nonumber\\
&&-D_\mathrm{0},\nonumber
\end{eqnarray}
\begin{eqnarray}
&&C_\mathrm{R_{B}}=-\Omega_\mathrm{R_{B}}\tanh^2\chi\left[1+\left(\Omega_\mathrm{T_{A}}-1\right)\tanh^2\chi\right]\left[1+\left(\Omega_\mathrm{R_{A}}-1\right)\tanh^2\chi\right] \left[1+\left(\Omega_\mathrm{T_{B}}-1\right)\tanh^2\chi\right]\\
&&+\left[1+\left(\Omega_\mathrm{T_{A}}-1\right)\tanh^2\chi\right]\left[1+\left(\Omega_\mathrm{T_{B}}-1\right)\tanh^2\chi\right]D_\mathrm{T_A,T_B}
+\Omega_\mathrm{R_{B}}\tanh^2\chi\left[1+\left(\Omega_\mathrm{T_{A}}-1\right)\tanh^2\chi\right]D_\mathrm{T_A,R_B}\nonumber\\
&&+\Omega_\mathrm{R_{B}}\tanh^2\chi\left[1+\left(\Omega_\mathrm{R_{A}}-1\right)\tanh^2\chi\right]D_\mathrm{R_A,R_B}
+\left[1+\left(\Omega_\mathrm{R_{A}}-1\right)\tanh^2\chi\right]\left[1+\left(\Omega_\mathrm{T_{B}}-1\right)\tanh^2\chi\right]D_\mathrm{R_A,T_B}\nonumber\\
&&-D_\mathrm{0},\nonumber
\end{eqnarray}
\begin{eqnarray}
&&C_\mathrm{T_{A},T_{B}}=\Omega_\mathrm{T_{A}}\Omega_\mathrm{T_{B}}\tanh^4\chi\left[1+\left(\Omega_\mathrm{R_{A}}-1\right)\tanh^2\chi\right]\left[1+\left(\Omega_\mathrm{R_{B}}-1\right)\tanh^2\chi\right]\\
&&-\Omega_\mathrm{T_{A}}\Omega_\mathrm{T_{B}}\tanh^4\chi
D_\mathrm{T_A,T_B}
-\Omega_\mathrm{T_{A}}\tanh^2\chi\left[1+\left(\Omega_\mathrm{R_{B}}-1\right)\tanh^2\chi\right]D_\mathrm{T_A,R_B}\nonumber\\
&&-\left[1+\left(\Omega_\mathrm{R_{A}}-1\right)\tanh^2\chi\right]\left[1+\left(\Omega_\mathrm{R_{B}}-1\right)\tanh^2\chi\right]D_\mathrm{R_A,R_B}
-\Omega_\mathrm{T_{B}}\tanh^2\chi\left[1+\left(\Omega_\mathrm{R_{A}}-1\right)\tanh^2\chi\right]D_\mathrm{R_A,T_B}\nonumber\\
&&+D_\mathrm{0},\nonumber
\end{eqnarray}
\begin{eqnarray}
&&C_\mathrm{T_{A},R_{B}}=\Omega_\mathrm{T_{A}}\Omega_\mathrm{R_{B}}\tanh^4\chi\left[1+\left(\Omega_\mathrm{R_{A}}-1\right)\tanh^2\chi\right] \left[1+\left(\Omega_\mathrm{T_{B}}-1\right)\tanh^2\chi\right]\\
&&-\Omega_\mathrm{T_{A}}\tanh^2\chi\left[1+\left(\Omega_\mathrm{T_{B}}-1\right)\tanh^2\chi\right]D_\mathrm{T_A,T_B}
-\Omega_\mathrm{T_{A}}\Omega_\mathrm{R_{B}}\tanh^4\chi D_\mathrm{T_A,R_B}\nonumber\\
&&-\Omega_\mathrm{R_{B}}\tanh^2\chi\left[1+\left(\Omega_\mathrm{R_{A}}-1\right)\tanh^2\chi\right]D_\mathrm{R_A,R_B}
-\left[1+\left(\Omega_\mathrm{R_{A}}-1\right)\tanh^2\chi\right]\left[1+\left(\Omega_\mathrm{T_{B}}-1\right)\tanh^2\chi\right]D_\mathrm{R_A,T_B}\nonumber\\
&&+D_\mathrm{0},\nonumber
\end{eqnarray}
\begin{eqnarray}
&&C_\mathrm{R_{A},T_{B}}=\Omega_\mathrm{R_{A}}\Omega_\mathrm{T_{B}}\tanh^4\chi\left[1+\left(\Omega_\mathrm{T_{A}}-1\right)\tanh^2\chi\right]\left[1+\left(\Omega_\mathrm{R_{B}}-1\right)\tanh^2\chi\right]\\
&&-\Omega_\mathrm{T_{B}}\tanh^2\chi\left[1+\left(\Omega_\mathrm{T_{A}}-1\right)\tanh^2\chi\right]D_\mathrm{T_A,T_B}
-\left[1+\left(\Omega_\mathrm{T_{A}}-1\right)\tanh^2\chi\right]\left[1+\left(\Omega_\mathrm{R_{B}}-1\right)\tanh^2\chi\right]D_\mathrm{T_A,R_B}\nonumber\\
&&-\Omega_\mathrm{R_{A}}\tanh^2\chi\left[1+\left(\Omega_\mathrm{R_{B}}-1\right)\tanh^2\chi\right]D_\mathrm{R_A,R_B}
-\Omega_\mathrm{R_{A}}\Omega_\mathrm{T_{B}}\tanh^4\chi D_\mathrm{R_A,T_B}\nonumber\\
&&+D_\mathrm{0},\nonumber
\end{eqnarray}
\begin{eqnarray}
&&C_\mathrm{R_{A},R_{B}}=\Omega_\mathrm{R_{A}}\Omega_\mathrm{R_{B}}\tanh^4\chi\left[1+\left(\Omega_\mathrm{T_{A}}-1\right)\tanh^2\chi\right] \left[1+\left(\Omega_\mathrm{T_{B}}-1\right)\tanh^2\chi\right]\\
&&-\left[1+\left(\Omega_\mathrm{T_{A}}-1\right)\tanh^2\chi\right]\left[1+\left(\Omega_\mathrm{T_{B}}-1\right)\tanh^2\chi\right]D_\mathrm{T_A,T_B}
-\Omega_\mathrm{R_{B}}\tanh^2\chi\left[1+\left(\Omega_\mathrm{T_{A}}-1\right)\tanh^2\chi\right]D_\mathrm{T_A,R_B}\nonumber\\
&&-\Omega_\mathrm{R_{A}}\Omega_\mathrm{R_{B}}\tanh^4\chi
D_\mathrm{R_A,R_B}
-\Omega_\mathrm{R_{A}}\tanh^2\chi\left[1+\left(\Omega_\mathrm{T_{B}}-1\right)\tanh^2\chi\right]D_\mathrm{R_A,T_B}\nonumber\\
&&+D_\mathrm{0},\nonumber
\end{eqnarray}
where
\begin{equation}
D_\mathrm{T_{A},T_{B}}=\Omega_{\mathrm{R_{A}}}\Omega_{\mathrm{R_{B}}}\tanh^2\chi
\left|e^{-i\varphi} \sin\theta_\mathrm{A}\cos\theta_\mathrm{B}+
\cos\theta_\mathrm{A}\sin\theta_\mathrm{B}\right|^2,
\end{equation}
\begin{equation}
D_\mathrm{T_{A},R_{B}}=\Omega_{\mathrm{R_{A}}}\Omega_{\mathrm{T_{B}}}\tanh^2\chi
\left|e^{-i\varphi} \sin\theta_\mathrm{A}\sin\theta_\mathrm{B}-
\cos\theta_\mathrm{A}\cos\theta_\mathrm{B}\right|^2,
\end{equation}
\begin{equation}
D_\mathrm{R_{A},R_{B}}=\Omega_{\mathrm{T_{A}}}\Omega_{\mathrm{T_{B}}}\tanh^2\chi
\left|-e^{-i\varphi} \cos\theta_\mathrm{A}\sin\theta_\mathrm{B}-
\sin\theta_\mathrm{A}\cos\theta_\mathrm{B}\right|^2,
\end{equation}
\begin{equation}
D_\mathrm{R_{A},T_{B}}=\Omega_{\mathrm{T_{A}}}\Omega_{\mathrm{R_{B}}}\tanh^2\chi
\left|-e^{-i\varphi}\cos\theta_\mathrm{A}\cos\theta_\mathrm{B}+
\sin\theta_\mathrm{A}\sin\theta_\mathrm{B}\right|^2,
\end{equation}
\begin{equation}
D_\mathrm{0}=\Omega_{\mathrm{T_{A}}}\Omega_{\mathrm{R_{A}}}\Omega_{\mathrm{T_{B}}}
\Omega_{\mathrm{R_{B}}}\tanh^4\chi,
\end{equation}
\begin{equation}
\Omega_\mathrm{T_{A(B)}}=\eta_{\scriptscriptstyle
\mathrm{T_{A(B)}}}\left|T_\mathrm{A(B)}\right|^{2},
\end{equation}
\begin{equation}
\Omega_\mathrm{R_{A(B)}}=\eta_{\scriptscriptstyle
\mathrm{R_{A(B)}}}\left|T_\mathrm{A(B)}\right|^{2}.
\end{equation}

\end{widetext}


\begin{thebibliography}{99}
\bibitem{BS} The experimental setup in Ref.~\cite{Fedrizzi} consists of an
additional beam-splitter at the receiver station. Here we do not
consider the corresponding field transformations. The related losses
of $3\,\mathrm{dB}$ are included in the detection losses.
\end{thebibliography}
\end{document}